\documentclass{amsart}
\usepackage{amssymb}
\usepackage[arrow,curve,all]{xy}
\def\intprod{\mathbin{\hbox to 6pt{%
                 \vrule height0.4pt width5pt depth0pt
                 \kern-.4pt
                 \vrule height6pt width0.4pt depth0pt\hss}}}

\let\hook\intprod
\newcommand{\mspym}{(\mathcal{Z},\Omega^\mathcal{Z})}
\newcommand{\vdjbym}{(J^1 (\pi^{\text{YM}})^*,\Omega^V_{\text{YM}})}
\newcommand{\cmspym}{(\mathcal{Z}^\mathfrak{T},\Omega^\mathfrak{T})}
\newcommand{\vcdjbym}{(J^1 (\pi_\mathfrak{T}^{\text{YM}})^*,\Omega^{V\mathfrak{T}}_{\text{YM}})}
\newcommand{\cdjbym}{(J^1 (\pi_\mathfrak{T}^{\text{YM}})^*,\Omega^{\text{YM}}_{\mathcal{H}^{\text{YM}}})}
\newcommand{\smspym}{(\mathcal{S}\pi^\mathbb{L},\Omega^\mathbb{L})}
\newcommand{\csmspym}{(\mathcal{S}\pi_\mathfrak{T}^\mathbb{L},\Omega^{\mathbb{L}\mathfrak{T}})}
\newcommand{\vsdjbym}{(J^1 (\pi^{\text{YM}}_{\mathbb{L}\mathfrak{T}})^*,\Omega_{\mathbb{L}\mathfrak{T}}^{V\text{YM}})}

\newcommand{\carb}{\Omega^V_{\text{YM}}}
\newcommand{\Tb}{\Theta^V_{\text{YM}}}
\newcommand{\carc}{\Omega^\mathfrak{T}}
\newcommand{\card}{\Omega^{V\mathfrak{T}}_{\text{YM}}}

\newcommand{\care}{\Omega^{\text{YM}}_{\mathcal{H}^{\text{YM}}}}

\newcommand{\carh}{\Omega_{\mathbb{L}\mathfrak{T}}^{V\text{YM}}}

\newcommand{\group}{$\mathfrak{G}$ }

\newcommand{\lalg}{$\mathfrak{g}$ }

\newcommand{\hftv}{\Tilde{\mathcal{H}}^*(M,\Omega)}

\newcommand{\dhftv}{\Tilde{\mathcal{H}}^*(J^1\pi^*,\Omega)}
\newcommand{\sdhftv}{\Tilde{\mathcal{H}}^*(J^{1}(\pi^ {\Pi\mathfrak{L}B})^*,\Omega^{\Pi\mathfrak{L}B}_\mathfrak{H})}

\newcommand{\djb}{J^{1} \pi^*}             
\newcommand{\jb}{J^1 \pi}                  
\newcommand{\sdjb}{J^{1}{\pi^{\Pi\mathfrak{L}B}}^*} 
\newcommand{\sdjbh}{(J^{1}(\pi^{\Pi\mathfrak{L}B})^*,\Omega^{\Pi\mathfrak{L}B}_\mathfrak{H})}
\newcommand{\sjb}{J^1 \pi^{\Pi\mathfrak{L}B}}    
\newcommand{\msp}{\mathcal{M}\pi}    
\newcommand{\smsp}{\mathcal{S}\pi}      
\newcommand{\lcb}{\pi^{\mathfrak{L}B}}
\newcommand{\lscb}{\pi^{\Pi\mathbb{L}B}}      
\newcommand{\ljb}{J^1 \pi^{{\mathfrak{L}}B}}   
\newcommand{\ldjb}{J^1 \pi^{\mathfrak{L}B *}}
\newcommand{\ldjbo}{J^1 \pi^{{\mathfrak{L}0} *}}
\newcommand{\lmsp}{\mathcal{M}\pi^{\mathfrak{L}B}} 
\newcommand{\lmspo}{\mathcal{M}\pi^{\mathfrak{L}0}} 
\newcommand{\lsjb}{J^1 \pi^{\Pi\mathbb{L}B}}    
\newcommand{\lsdjb}{J^1 \pi^{ \Pi\mathbb{L}B *}} 
\newcommand{\lsdjbh}{(J^1 \pi^{ \Pi\mathbb{L}B *},\Omega^{\Pi\mathbb{L}B}_\mathfrak{H})}
\newcommand{\lsmsp}{\mathcal{S}\pi^{\mathbb{L}}}

\newcommand{\isto}{\hspace{5pt} ::\hspace{5pt}} 
\newtheorem{thm}{Theorem}[section]

\newtheorem{cor}[thm]{Corollary}
\newtheorem{prop}[thm]{Proposition}
\newtheorem{defn}[thm]{Definition}

\begin{document}
\title[On a Multisymplectic Formulation of the Classical BRST Symmetry...]
    {On a Multisymplectic Formulation of the Classical BRST Symmetry for First Order Field Theories Part II: Geometric Structures.
    }
\author{S.P.Hrabak.}
\address{Department of Mathematics\\
  King's College London\\ 
  Strand\\
   London WC2R 2LS\\
 England} 
\email{e-mail: shrabak@mth.kcl.ac.uk}
\thanks{Research supported by PPARC}

\subjclass{Primary: 53  Secondary: 70}
\date{January 20, 1999}

\begin{abstract}
A geometric multisymplectic formulation of the classical BRST symmetry of constrained first-order classical field theories is  described. To effect this we introduce graded analogues of the bundles and manifolds of the multisymplectic formulation of first-order field theories. The Lagrange-d'Alembert formalism is also developed in terms of the  multisymplectic framework. The result is a covariant Hamiltonian BFV formalism.
\end{abstract}

\maketitle

In an accompanying preceding paper \cite{jdg1} the author detailed the homological algebra which provided an algebraic description of the Marsden-Weinstein multisymplectic reduction. The study of multisymplectic geometry \cite{Cans,CanG,CanH,Ibort,Martin} arose in the context of the search for the geometric foundations of classical field theory \cite{Kij1,Kij2,KijSzc1,KijSzc2}. We turn in this paper to a study of those multisymplectic manifolds which form the geometric foundations of first order classical field theories. We develop a geometric multisymplectic formalism describing the classical BRST\footnote{The classical BRST symmetry is a Grassmann-odd symmetry (symplectomorphism) which was originally introduced within the framework of the quantisation of constrained dynamical systems, only later was its relation to reduction in the classical context understood \cite{KS,HennTit}.} symmetry of constrained first-order classical field theories whose symmetries arise by virtue of the prolongation of a group action by bundle morphisms of the configuration bundle. The essence of the  geometric formulation is to encode the algebraic 
structures of the homological description of Marsden-Weinstein multisymplectic reduction  into  geometric structures on graded-multisymplectic manifolds.

\begin{enumerate}
\item In the first section we introduce the Lagrange multiplier into the multisymplectic framework. We effect this by virtue of the introduction of a certain integrable distribution whose existence follows by virtue of the assumed free and proper group action by bundle automorphisms on the configuration bundle. We make use of this  distribution  in order to define a configuration bundle,  multiphase space and covariant phase space extended by the addition of a Lagrange multiplier and its canonical momenta. We then generalise the  Lagrange-d'Alembert formalism to the multisymplectic context whence we obtain {\it both} the generalised Lagrange-d'Alembert-Hamiltonian equations of motion and the conservation of the covariant Noether currents from a single elegant geometric equation.
\item In the second section  we construct  graded analogues of the configuration bundle, the  multiphase space and the corresponding  covariant phase space, of the multisymplectic formalism. The geometric constructions described here formalise the introduction of the Grassmann-odd degrees of freedom known in the literature as ghosts. 
\item In the third section we shall complete the translation of the algebraic 
structures needed in the homological description of Marsden-Weinstein multisymplectic reduction into  geometric structures on graded multisymplectic manifolds. We introduce a certain Abelian Grassmann-odd bundle morphism of the graded configuration bundle which upon prolongation to the graded covariant phase space is the classical BRST symmetry. Upon   projection of the Abelian Grassmann-odd bundle morphism to the configuration-bundle one regains the original non-Abelian gauge automorphism of the configuration bundle. The observables on the graded multisymplectic manifold form a $\mathbb{Z}_2\times\mathbb{Z}_2$-graded Poisson-Leibniz algebra. We identify the algebraic differential complex of \cite{jdg1} with a pair consisting of a subalgebra of the  algebra of observables and the Poisson-Leibniz derivation generated by a  Grassmann-odd (n-1)-form. The Grassmann-odd (n-1)-form is the lifted momentum observable corresponding to the  Abelian Grassmann-odd bundle morphism of the graded configuration bundle. The observables on the reduced multisymplectic manifold are obtained as the zeroth homology of this geometric differential complex.
\item In the fourth  section we combine the Lagrange d'Alembert formalism with the graded multisymplectic formalism in order to obtain a covariant Hamiltonian BFV formalism.
\item In the fifth section we shall illustrate the new formalism by deriving a new BRST algebra for the well known  exemplar of Yang and Mills. The novelty in our covariant Hamiltonian approach is that the resulting BRST algebra is  polynomial in the canonical variables, unlike the non-covariant approach where one finds the algebra to include spatial derivatives of the canonical variables.
\item We conclude in the final section with an appendix containing a brief description of those elements of the multisymplectic formulation of first order field theories prerequisite to the exposition in this paper.
\end{enumerate}
We  thus  obtain a covariant Hamiltonian geometric formulation of the classical BRST symmetry for first order field theories.
				
\section{A Multisymplectic  Lagrange-d'Alembert Formalism.}

The BFV formulation \cite{BFV} of the classical BRST symmetry includes a pair of dynamical variables which are  the Lagrange multiplier of the Lagrange-d'Alembert variational principle and  canonically conjugate momenta which is prescribed to vanish so that the introduction of the Lagrange multiplier does not introduce extra dynamical degrees of freedom. Since we should like to construct  a multisymplectic formulation of the classical BRST symmetry it is therefore necessary to formulate the existence of this  pair within the multisymplectic formalism. In the first subsection we shall introduce the Lagrange multiplier distribution. We thus encode the notion of a Lagrange multiplier into a geometric structure. We shall show that our characterisation generalises the usual definition in the symplectic to the multisymplectic case. In the second subsection we shall use the Lagrange multiplier distribution in order to define a configuration bundle and multiphase space extended by the addition of a Lagrange multiplier and a canonically conjugate multimomenta. In the final subsection  we shall formulate a generalisation of the Hamiltonian equations of the Lagrange-d'Alembert variational principle for field theories in terms of a geometric equation on a certain multisymplectic manifold.

\subsection{The Lagrange Multiplier Distribution}
In order to formulate the Hamiltonian BFV structures within the multisymplectic formalism we need to introduce Lagrange multipliers and the canonically conjugate momenta. We obviously need to know how to capture within the multisymplectic framework the notion of the Lagrange multiplier. A suitable starting point is an intrinsic property of the Lagrange multiplier within classical mechanics, namely that the Lagrange multiplier is a section of the cotangent bundle that annihilates the constraint distribution. In the following we shall extend this characterisation to the case of those free and proper group actions on  multisymplectic manifolds that arise by virtue of the prolongation of automorphisms of $E$ covering  diffeomorphisms of $B$ (viz. a morphism of the configuration bundle  see \S6). All gauge symmetries are of this type. 

Let $\{\xi_a\}_{a=1,\cdots,dim\mathfrak{G}}$ be a basis of the Lie Algebra $\mathfrak{g}$. Then if $\mathfrak{G}$ acts by automorphisms of $E$ covering diffeomorphisms of $B$, then the corresponding vector fields are projectable, namely in the adapted local coordinates they take the form 
$ \xi^E_a = \xi_a^{\alpha}(x) \frac{\partial}{\partial x^{\alpha}}
+\xi_a^i(x,u) \frac{\partial}{\partial u^i}$.
 The $\xi^E_a(x,u)$ span a subspace of the tangent space of $E$ at $(x,u)$. Let $\mathfrak{L}_{(x,u)}E:= <\xi^E_a(x,u)>$. Since we assume that the action of $\mathfrak{G}$ is free, $dim\mathfrak{L}_{(x,u)}E=dim\mathfrak{G} \hspace{5pt} \forall \hspace{5pt} (x,u) \varepsilon E$. We therefore have a distribution over $E$, namely $\mathfrak{L}E:= \underset{(x,u) \varepsilon E}{\sqcup}\mathfrak{L}_{(x,u)}E$.

{\defn Let $\pi^\mathfrak{L}$ denote the {\it Lagrange multiplier distribution} $(\mathfrak{L}E,\pi^\mathfrak{L},E)$.}

Having defined the distribution $\pi^\mathfrak{L}$ we need to determine the local adapted coordinates on $\pi^\mathfrak{L}$ induced from those on $\pi$. Recall that the local adapted coordinates of a point $p\varepsilon\mathcal{U}\subset E$ are $(x^\alpha,u^i)$ (see \S6). By definition the fibres of the distribution $\pi^\mathfrak{L}$ have a global basis given by $\{\xi^E_a\}_{a=1,\cdots,dim\mathfrak{G}}$. Any element $\mathfrak{v}\varepsilon\mathfrak{L}_pE$ of the fibre at p may therefore be written in the form $\mathfrak{v}=\underset{a=1,\cdots,dim\mathfrak{G}}{\Sigma}\lambda^a{\xi^E_a}$. The $\lambda^a$ are uniquely determined by $(x^\alpha,u^i)$ and $\mathfrak{v}$. They therefore determine coordinate functions $\lambda:\mathfrak{L}_pE\longrightarrow\mathbb{R}^{dim\mathfrak{G}}$. In summary we have a coordinate chart $\mathcal{U}\times\mathfrak{L}_pE\longrightarrow\mathbb{R}^{n+m}\times\mathbb{R}^{dim\mathfrak{G}}\isto (p,\mathfrak{v})\longrightarrow (x^\alpha,u^i,\lambda^a)$. In \cite{BFV} the Lagrange multipliers are introduced as functions which ``parametrise'' the gauge orbits. In the geometric formulation presented here this is made more precise by the explicit introduction of the distribution $\pi^\mathfrak{L}$.

{\prop The Lagrange Multiplier Distribution $\pi^\mathfrak{L}$ is {\it integrable}.}

\begin{proof}  Let $X,Y\varepsilon\mathfrak{L}_{(x,u)}$, then in terms of local coordinates they may be written as $X=X^a\xi^E_a$ and $Y=Y^a\xi^E_a$. So that $[X,Y]=[X^a\xi^E_a,Y^b\xi^E_b]=X^aY^b[\xi^E_a,\xi^E_b]+(X^a\xi^E_aY^c-Y^b\xi^E_bX^c)\xi^E_c=(X^aY^bC_{ab}^c+X^a\xi^E_aY^c-Y^b\xi^E_bX^c)\xi^E_c\varepsilon\mathfrak{L}_{(x,u)}$ \end{proof}

{\defn The {\it constraint distribution} $\mathcal{D}$ is the tangent bundle of the {\it constraint submanifold} , namely $T\mathcal{C}$.} 

In the context of interest to us, the constraint submanifold is the manifold defined as the inverse image of the  covariant momentum map corresponding to the lift of the $\mathfrak{G}$-action by bundle morphisms on the configuration bundle $\pi$ to the multiphase space $\msp$. It is a distribution as the dimension of each fibre is constant by the assumed  regularity of the covariant momentum map . From the Marsden-Weinstein multisymplectic reduction theorem (see \cite{jdg1}) we learn that the Hamiltonian vector fields corresponding to the covariant Noether currents span the symplectic complement of the constraint distribution. The  prolongations of the $\xi^E_a$ to the multiphase space thus span the symplectic complement of the constraint distribution. The property that the Lagrange multiplier is the annihilator of the constraint distribution is thus encoded within the following result.
 
{\prop The prolongation of the automorphism of $\pi$ generated by the $\xi^E_a$ spans a distribution over $\msp$ that is the {\it annihilator} of the {\it constraint distribution} $\mathcal{D}$.}

\subsection{The Lagrange Multiplier Extended Bundles}

This subsection is an exposition of the construction of the extended bundles associated with the introduction of Lagrange multipliers into the multisymplectic formulation of first order field theories.
The Lagrange multiplier distribution $\pi^\mathcal{L}$ defines a bundle $\pi^{\mathcal{L}B}$ over B by virtue of the first of the two diagrams below. The projection $\pi^{\mathcal{L}B}$ is thus defined to be $\pi^{\mathcal{L}B}:=\pi \circ \pi^\mathcal{L}$.
\[
\xymatrix{
\mathfrak{L}E \ar[r]^{\pi^{\mathfrak{L}}} \ar[d]_{\pi^{\mathcal{L}B}}& E \ar[dl]^\pi & & \ljb \ar[d]_{\pi^{0 \mathfrak{L}B}} \ar[r]^{\pi^{0 \mathfrak{L}B}_1} & \ar[dl]^{\pi^{\mathfrak{L}B}} \mathfrak{L}B\\
B & & &  B\\
}
\] 
{\defn The {\it extended configuration bundle} $\pi^{\mathcal{L}B}$ is the triple $(\mathfrak{L}E,\pi^{{\mathcal{L}B}},B)$.}

The bundle $\pi^{\mathcal{L}B}$ has local adapted coordinates $(x^{\alpha},u^i,\lambda^a)$. Although $\pi^{\mathfrak{L}}$ and $\pi^{\mathcal{L}B}$ have the same local adapted coordinates an important difference lies in  the structures of their respective vertical bundles. Recall that the vertical bundle plays an important role in the definition of the multiphase space (see \S6). The structure of the vertical bundle $V\pi^{\mathcal{L}B}$ leads to a   multiphase space extended by the additional structure of  a Lagrange multiplier and the canonically conjugate multi-momenta. The first jet bundle $\ljb$ over $\pi^{\mathcal{L}B}$ has one jets which in local coordinates take the form $j^1_{(u,\lambda)}\phi(x)=(x^{\alpha},u^i,\lambda^a,u^i_{,{\alpha}},\lambda^a_{,{\alpha}})$. Note that in comparison  the one jets of the first jet bundle over $\pi^{\mathfrak{L}}$ take the local form $j^1_{(u,\lambda)}\phi(x)=(x^{\alpha},u^i,\lambda^a,u^i_{,{\alpha}},\lambda^a_{,{\alpha}},u^i_{,j},\lambda^a_{,j})$. The first jet bundle over $\pi^{\mathfrak{L}}$ is therefore a much larger bundle than we want. The commutivity of the second of the two diagrams above  defines  the surjective submersion $\pi^{0 \mathfrak{L}B}:= \pi^{\mathfrak{L}B}   \circ     \pi^{0 \mathfrak{L}B}_1$.  We define the affine dual of the first jet bundle over $\pi^{\mathcal{L}B}$ by the following short exact sequence (c.f. \S6):
\[
\xymatrix{
0 \ar[r] & \Lambda^n_0(\lcb) \ar[r]^i& \Lambda^n_1 (\lcb) \ar[r]^{\varrho^{\mathfrak{L}B}} & \ldjb \ar[r] & 0
}
\]
{\prop
$\Lambda^n_1 (\pi^{\mathcal{L}B})$ is endowed with a canonical n-form $\Theta^{\mathfrak{L}B}$. In local coordinates it takes the form $\Theta^{\mathfrak{L}B}=pd^n x + p_i^{\alpha}du^i \wedge d^{n-1}x_{\alpha}
+B^{\alpha}_a d\lambda^a \wedge d^{n-1}x_{\alpha}$.

}
\begin{proof}Let $\omega$ denote an  n-form on $\lcb$ and let $\Tilde{\omega}$ denote the corresponding section. They are related by a canonical n-form $\Theta{^\mathfrak{L}B}$ on $ \lcb$ according to $\Tilde{\omega}^*\Theta^{\mathfrak{L}B}=\omega$. The point $\omega$ in $\Lambda^n_1 (\lcb)$ with coordinates $(x^{\alpha},u^i,\lambda^a,p_i^{\alpha},B_a^{\alpha},p)$ is the n-form $\omega=pd^n x + p_i^{\alpha}du^i \wedge d^{n-1}x_{\alpha}
+B^{\alpha}_a d\lambda^a \wedge d^{n-1}x_{\alpha}$ at $(x^{\alpha},u^i,\lambda^a)$ of $\pi^{\mathcal{L}B}$. Then $\Theta^{\mathfrak{L}B}$ is uniquely determined to be 
$$\Theta^{\mathfrak{L}B}=pd^n x + p_i^{\alpha}du^i \wedge d^{n-1}x_{\alpha}
+B^{\alpha}_a d\lambda^a \wedge d^{n-1}x_{\alpha}$$
by the tautology $\Tilde{\omega}^*\Theta^{\mathfrak{L}B}=\omega$.\end{proof}
{\defn
Let $\mathcal{M}\pi^{\mathfrak{L}B}:=\Lambda^n_1(\pi^{\mathcal{L}B}) $. We call the pair $(\mathcal{M}\pi^{\mathfrak{L}B},\Omega^{\mathfrak{L}B})$, where $\Omega^{\mathfrak{L}B}= -d\Theta^{\mathfrak{L}B}$  the {\it  extended multiphase space }. 
}

In terms of the local adapted coordinates on $\mathcal{M}\pi^{\mathfrak{L}B}$, namely $(x^{\alpha},u^i,\lambda^a,p_i^{\alpha},B_a^{\alpha},p)$, the Cartan (n+1)-form takes the form 
\begin{equation}\Omega^{\mathfrak{L}B}=-dp\wedge d^n x +du^i \wedge dp_i^{\alpha} \wedge d^{n-1}x_{\alpha}
+d\lambda^a\wedge dB^{\alpha}_a  \wedge d^{n-1}x_{\alpha}.
\end{equation}
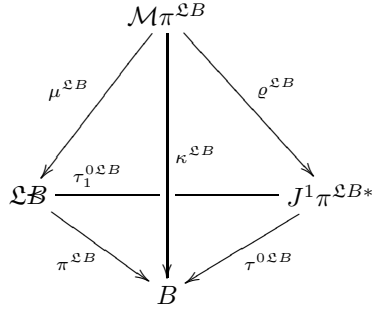
\[
\xymatrix{
&\lmsp \ar[ddl]_{\mu^{\mathfrak{L}B}} \ar[ddr]^{\varrho^{\mathfrak{L}B}} \ar[ddd]^{\kappa^{\mathfrak{L}B}} \\
&& \\
\mathfrak{L}B \ar[dr]_{\pi^{\mathfrak{L}B}}  && \ldjb \ar''[l]'[ll]_{\tau^{0 \mathfrak{L}B }_1} \ar[dl]^{\tau^{0 \mathfrak{L}B}} \\
&B }
\]
The above commuting diagram defines the  surjective submersions:
$\mu^{\mathfrak{L}B} := \tau^{0 \mathfrak{L}B }_1 \circ \varrho^{\mathfrak{L}B}	; \quad \kappa^{\mathfrak{L}B}:= \pi^{\mathfrak{L}B} \circ \mu^{\mathfrak{L}B}; \quad \tau^{0 \mathfrak{L}B }:=\pi^{\mathfrak{L}B} \circ \tau^{0 \mathfrak{L}B }_1.$

\subsection{The Covariant Hamiltonian Equations of the Lagrange-d'Alembert Formalism}
For a classical mechanical system subject to constraints there exists a formulation of the principle of least action called the Lagrange-d'Alembert principle by virtue of which  the Lagrangian equations of motion on the constraint submanifold are obtained. For our purposes, the corresponding Hamiltonian equations are  more interesting for in this section we shall formulate the covariant analogue of Hamilton's equations in a  Lagrange-d'Alembert formalism for first order field theories. 

For classical mechanical systems with generalised coordinates $q^i$, generalised momenta $p_i$ ,where $i=1,\cdots, m$ , Hamiltonian $\mathcal{H}$ subject to constraints $\phi^a$, where $a=1,\cdots ,dim\mathfrak{G}$, the Lagrange-d'Alembert-Hamilton equations take the following form 
\begin{equation}
\begin{split}
\dot{q^i}&=\frac{\partial \mathcal{H}}{\partial p_i } + u_m\frac{\partial\phi^m }{\partial p_i  } \\
-\dot{p_i}&=\frac{\partial \mathcal{H}}{\partial q^i  } + u_m\frac{\partial\phi^m }{\partial q^i }
\end{split}
\end{equation}
In \cite{Dirac,HennTit} these equations are obtained by virtue of a variational argument. As is characteristic of the multisymplectic formalism we shall find that the analogue of these equations for first order field theories may be written as a geometric equation.

We begin by considering the extended multiphase space $\lmsp$. Recall that the local adapted coordinates on $\lmsp$ are $(x^{\alpha},u^i,\lambda^a,p_i^{\alpha},B_a^{\alpha},p)$. Let $\lmspo$ be the submanifold of $\lmsp$ which is defined locally by the vanishing of the Lagrange multiplier momentum, namely $B_a^{\alpha}\approx0$. This is admissible because the triviality of the distribution  ${\mathfrak{L}E}$. Since the fibres ${\mathfrak{L}E}$ have a global basis  
${\mathfrak{L}E}\cong E\times\mathbb{R}^\mathfrak{G}$. It therefore follows that  $T{\mathfrak{L}E}\cong TE\times T\mathbb{R}^\mathfrak{G}$. The global coordinates $\{\lambda^a\}_{a=1,\cdots ,dim\mathfrak{G}}$ on $\mathbb{R}^\mathfrak{G}$ then define a canonical global basis of $T\mathbb{R}^\mathfrak{G}$, namely $\{\frac{\partial}{\partial\lambda^a}\}_{a=1,\cdots ,dim\mathfrak{G}}$. The lifted momentum observables based on the vector fields $\{\frac{\partial}{\partial\lambda^a}\}_{a=1,\cdots ,dim\mathfrak{G}}$ are then also globally defined. They take  the form $B_a:=B_a^\alpha d^{n-1}x_\alpha$. Declaring $B_a\approx 0$ then implies that $B_a^\alpha\approx 0$ globally, as the $\{d^{n-1}x_\alpha\}_{\alpha = 1, \cdots, n}$ are linearly independent. One may interpret the submanifold $\lmspo$ as the multiphase space of non-propagating Lagrange multipliers. 
Recall that in the formulation of the Lagrange-d'Alembert variational principle in mechanics we require that the Lagrange multipliers enforce the constraint but do no work, that is they have no propagating degrees of freedom. The selection of this submanifold is an encoding of this requirement.

$\lmspo$ has adapted local coordinates $(x^{\alpha},u^i,\lambda^a,p_i^{\alpha},p)$. Let $\mathfrak{i}$ be the immersion of $\lmspo$ into $\lmsp$. Then we may use this immersion to pull back the Cartan  n-form on $\lmsp$ to $\lmspo$. The result may be expressed in terms of the local adapted coordinates as $\Theta^0:= \mathfrak{i}^*\Omega^{\mathfrak{L}B}=p d^n x + p_i^{\alpha}du^i\wedge d^{n-1} x_{\alpha}$. Corresponding to the multiphase space $\lmspo$ there exists a covariant phase space $\ldjbo$ which we define  by a short exact sequence, namely
\[
\xymatrix{
0 \ar[r] & \Lambda^n_0(\pi^{\mathfrak{L}0}) \ar[r]^i& \lmspo \ar[r]^{\varrho^{\mathfrak{L}}} & \ldjbo \ar[r] & 0
}
\]
Recall that the covariant Hamiltonian equations are encoded entirely within the the Cartan (n+1)-form. Clearly if we wish to obtain the Hamiltonian equations   of the Lagrange-d'Alembert formalism we must enrich the structure of the Cartan form. We achieve this end by adding to the Cartan n-form a term of the form $-\lambda^a d \delta(\xi_a)$, where we write $\delta(\xi_a)$ for the pull-back $\mathfrak{i}^*\delta(\xi_a)$.

{\defn The {\it extended Cartan (n+1)-form} is \[\Omega^{EX}:=-d(\Theta^0 - \lambda^a d \delta(\xi_a) )\] In the local adapted coordinates it takes the form  \begin{equation}\Omega^{EX}=-dp\wedge d^n x -dp_i^{\alpha}\wedge du^i\wedge d^{n-1} x_{\alpha}+ d\lambda^a\wedge  d \delta(\xi_a).\end{equation}}

Let $\mathfrak{H}$ be the Hamiltonian defined by $ p=-\mathfrak{H}$, by virtue of the Legendre transformation \cite{cramnotes,Crampin}. We pull back the extended Cartan form from $\lmspo$ to $\ldjbo$ by virtue of the section $\mathfrak{p}$ determined by $ p=-\mathfrak{H}$. Let $\Omega^{EX}_\mathfrak{H}:=\mathfrak{p}^*\Omega^{EX}$. In the local adapted coordinates $\Omega^{EX}_\mathfrak{H}$ takes the form $\Omega^{EX}_\mathfrak{H}=d\mathfrak{H}\wedge d^n x + du^i\wedge dp_i^{\alpha}\wedge  d^{n-1} x_{\alpha}+ d\lambda^a\wedge  d \delta_\mathfrak{H}(\xi_a)$, where $\delta_\mathfrak{H}(\xi_a)=\mathfrak{p}^*\delta(\xi_a)=(p_i^{\alpha}\xi_a^i -\mathfrak{H}\xi_a^{\alpha})d^{n-1}x_{\alpha} - p_i^{\alpha}\xi_a^{\beta}du^i \wedge
\partial_{\beta} \hook (\partial_{\alpha} \hook d^n x)$ \cite{MarShk}. 

{\defn Let $\Check{\phi}$ be the prolongation to $\lmspo$ of a section $\phi$ of $\pi^{\mathfrak{L}B}$ . $\Check{\phi}$ satisfies the covariant Lagrange-d'Alembert-Hamilton equations iff \begin{equation}\Check{\phi}^*(\mathcal{U}\hook\Omega^{EX}_\mathfrak{H})=0\quad \forall\text{ }\mathcal{U}\varepsilon \Gamma (\ldjbo,T\ldjbo).\end{equation}}

{\prop Let $G_a^{\alpha}:=p_i^{\alpha}\xi_a^i$ and $F^{ \beta{\alpha}}_{ai}:=p_i^{\alpha}\xi^\beta_a$. For the case of the generators $\delta(\xi_a)$ arising by virtue of an automorphism of $E$ covering the non-trivial diffeomorphisms  on $B$ then in terms of the local adapted coordinates the covariant Lagrange-d'Alembert-Hamiltonian equations take the  following form
\begin{equation}
\begin{split}
p_{i,{\alpha}}^{\alpha}&=-\frac{\partial\mathfrak{H}}{\partial u^i} + \frac{\partial G_a^{\alpha}}{\partial u^i}\lambda^a_{,{\alpha}} +\lambda^a_{,{\alpha}}\partial_\beta(F^{ \beta{\alpha}}_{ai}) -\lambda^a_{,{\beta}}\partial_\alpha(F^{ \beta{\alpha}}_{ai}) -\lambda^a_{,{\beta}}\xi^\beta_a\frac{\partial\mathfrak{H}}{\partial u^i}\\
u^i_{,\alpha}&=\frac{\partial\mathfrak{H}}{\partial p_i^{\alpha}} -\lambda^a_{,\beta}\frac{\partial G_a^{\beta}}{\partial p_i^{\alpha}}   +\lambda^a_{,\beta}\xi_a^\beta\frac{\partial\mathfrak{H}}{\partial p_i^{\alpha}} +\lambda^a_{,\alpha}\frac{\partial F^{ \beta{\gamma}}_{ak}}{\partial p_k^{\gamma}}u^i_{,{\beta}} -\lambda^a_{,\beta} \frac{\partial F^{ \beta{\gamma}}_{ak}}{\partial p_k^{\gamma}}u^i_{,{\alpha}}\\
0&=d\Check{\phi}^*\delta_\mathfrak{H}(\xi_a)\quad \text{ where here } d:=dx^\alpha \frac{\partial}{\partial x^\alpha}
\end{split}
\end{equation}}
\begin{proof}
It the local adapted coordinates $\Omega^{EX}_\mathfrak{H}$ takes the form:
\begin{equation}
\begin{split}
\Omega^{EX}_\mathfrak{H}&=\frac{\partial\mathfrak{H}}{\partial p_i^{\alpha}}dp_i^{\alpha}\wedge d^n x 
+\frac{\partial\mathfrak{H}}{\partial u^i} du^i \wedge d^n x \\
&+ du^i\wedge dp_i^{\alpha}\wedge  d^{n-1} x_{\alpha}
+ \xi^i_a d\lambda^a\wedge dp_i^{\alpha}\wedge  d^{n-1} x_{\alpha} \\
&+ \xi^i_{a,{\alpha}}p_i^{\alpha}d\lambda^a\wedge d^n x
+ \xi^i_{a,{j}}p_i^{\alpha}d\lambda^a\wedge du^i\wedge  d^{n-1} x_{\alpha} \\
&-\frac{\partial\mathfrak{H}}{\partial x^\alpha}\xi^\alpha_ad\lambda^a\wedge d^n x
-\frac{\partial\mathfrak{H}}{\partial u^i}\xi^\alpha_a d\lambda^a\wedge du^i\wedge  d^{n-1} x_{\alpha} \\
&-\frac{\partial\mathfrak{H}}{\partial p_k^{\beta}}\xi^\alpha_ad\lambda^a\wedge
dp_k^{\beta}\wedge  d^{n-1} x_{\alpha}
-\xi^\alpha_{a,{\alpha}}\mathfrak{H}d\lambda^a\wedge d^n x \\
&-\xi^\beta_a d\lambda^a\wedge dp_i^{\alpha}\wedge du^i \wedge d^{n-2} x_{\alpha\beta}
+ p_i^{\alpha}\xi^\beta_{a,{\gamma}}d\lambda^a\wedge du^i\wedge dx^\gamma \wedge d^{n-2} x_{\alpha\beta}
\end{split}
\end{equation}
Then
\begin{equation}
\begin{split}
\frac{\partial}{\partial u^l}\hook\Omega^{EX}_\mathfrak{H}&=\frac{\partial\mathfrak{H}}{\partial u^l}d^n x 
+dp_l^{\alpha}\wedge  d^{n-1} x_{\alpha} 
-\xi^i_{a,{l}}p_i^{\alpha}d\lambda^a\wedge d^{n-1} x_\alpha \\
&+\frac{\partial\mathfrak{H}}{\partial u^l}\xi^\alpha_a d\lambda^a\wedge  d^{n-1} x_{\alpha} 
-\xi^\beta_a d\lambda^a\wedge dp_l^{\alpha} \wedge d^{n-2} x_{\alpha\beta} \\
&-p_l^{\alpha}\xi^\beta_{a,{\beta}}d\lambda^a\wedge dx^\gamma \wedge d^{n-2} x_{\alpha\alpha} 
+p_l^{\alpha}\xi^\beta_{a,{\alpha}}d\lambda^a\wedge dx^\gamma \wedge d^{n-2} x_{\alpha\beta}
\end{split}
\end{equation}
\begin{equation}
\begin{split}
\frac{\partial}{\partial p_l^{\epsilon} }\hook\Omega^{EX}_\mathfrak{H}&=
\frac{\partial\mathfrak{H}}{\partial p_l^{\epsilon}}d^n x 
-du^l \wedge d^{n-1} x_{\epsilon} 
-\xi^l_ad\lambda^a\wedge d^{n-1} x_{\epsilon} \\
&+\frac{\partial\mathfrak{H}}{\partial p_l^{\epsilon}}\xi^\alpha_ad\lambda^a\wedge d^{n-1} x_{\alpha}
+ \xi^\beta_a d\lambda^a\wedge du^l\wedge  d^{n-2} x_{\epsilon \beta}
\end{split}
\end{equation}
from which follow Hamilton's equations of the  Lagrange-d'Alembert formalism.
\end{proof}
Then setting $\xi^\alpha_a$ equal to zero we obtain:
{\cor For the case of the generators $\delta(\xi_a)$ arising by virtue of an automorphism of $E$ covering the identity on $B$ then in terms of the local adapted coordinates the covariant Lagrange-d'Alembert-Hamiltonian equations take the form
\begin{equation}
\begin{split}
p_{i,{\alpha}}^{\alpha}&=-\frac{\partial\mathfrak{H}}{\partial u^i} + \frac{\partial G_a^{\alpha}}{\partial u^i}\lambda^a_{,{\alpha}} \\
u^i_{,\alpha}&=\frac{\partial\mathfrak{H}}{\partial p_i^{\alpha}} -\frac{\partial G_a^{\beta}}{\partial p_i^{\alpha}} \lambda^a_{,\beta} \\
0&=d\Check{\phi}^*\delta_\mathfrak{H}(\xi_a)\quad \text{ where here } d:=dx^\alpha \frac{\partial}{\partial x^\alpha}
\end{split}
\end{equation}}
The equation $0=d\Check{\phi}^*\delta_\mathfrak{H}(\xi_a)$ is the covariant form of {\it Noether's Theorem} (see \cite{MarShk}) from which we learn that the covariant Noether current $\Check{\phi}^*\delta_\mathfrak{H}(\xi_a)$ is constant, i.e. $dx^\alpha\partial_\alpha \Check{\phi}^*\delta_\mathfrak{H}(\xi_a)=0$. We therefore have obtained from our multisymplectic formulation of the Lagrange-d'Alembert principle both the Lagrange-d'Alembert-Hamiltonian equations of motion and the conservation of the Noether current. 
Note that when $dimB=1$ and upon identification  $u^a=\dot{\lambda}^a$ we regain the Lagrange-d'Alembert-Hamiltonian equations for classical mechanics.

\section{The Graded Multisymplectic Formalism}

In the usual formulation of the classical BRST symmetry the classical phase space is enlarged by introducing a Grassmann-odd coordinate, whose components equal in number the number of constraint equations, and  canonically conjugate momenta. These Grassmann-odd coordinates were first introduced  in the context of a path integral quantisation of constrained classical field theories. In that context they were termed ``ghosts'' since they were deemed to be unobservable. The BRST symmetry is then introduced as a Grassmann-odd symplectomorphism of this extended phase space. In this section we shall give a new formulation of these classical structures in the context of multisymplectic geometry. 

In this section we construct the graded configuration bundle, the graded multiphase space and the corresponding graded covariant phase space for which the BRST formalism takes the form of a Grassmann-odd, nilpotent multisymplectomorphism. We do not at this stage introduce the Lagrange multiplier. The formulation expounded in this section therefore corresponds to the minimal formulation of the Hamiltonian BFV where one introduces only ghosts and not the anti-ghosts \cite{HennTit}. In a section which is to follow we will construct  the Lagrange multiplier extension which  includes the introduction of anti-ghosts.

\subsection{The Graded Configuration Bundle}

We begin by defining the graded extension to the configuration bundle obtained by introducing the ghosts. We effect this extension by virtue of an elaboration of the Grassmann-even geometry of the configuration bundle $E$ to a $\mathbb{Z}_2$-graded geometry. Recall the functor $\varsigma$ from the category of smooth vector bundles $\mathfrak{V}$ to the category of graded-manifolds $\mathcal{S}(\mathfrak{V})$ associating the bundle with its total space obtained by reversing the Grassmann-parity of its  fibres. That is, $\mathfrak{V} \overset{\varsigma}{\longleftrightarrow} \mathcal{S}(\mathfrak{V}) \isto V\overset{\varsigma}{\longleftrightarrow} \Pi V$. $\Pi$ is the {\it Inversion Functor}, it is idempotent, and acts upon the fibres of vector bundles as $\Pi:\mathbb{R}^p\cong \mathbb{R}^{p|0} \rightarrow \mathbb{R}^{0|p}$, see \cite{voronov} for details.

In the definition which is to follow we shall use again the Lagrange multiplier distribution. This time we shall use the inversion functor in order to enlarge the configuration bundle by a Grassmann-odd superstructure. The appropriateness of this definition will become apparent as the ensuing structures unfold.
\[
\xymatrix{
\mathfrak{L}E \ar[rr]^\Pi \ar[d]_{\pi^{\mathfrak{L}}}& &\Pi\mathfrak{L}E \ar[d]^{\pi^{\Pi\mathfrak{L}}} \ar[ddl]_{\pi^{\Pi\mathfrak{L}B}}\\
E\ar[dr]_\pi& &E\ar[dl]^\pi \\
&B
}
\]
The above diagram illustrates the construction and defines the surjections $\pi^{\Pi\mathfrak{L}}:=\pi^{\mathfrak{L}} \circ \Pi$ and $\pi^{\Pi\mathfrak{L}B}:=\pi \circ \pi^{\Pi\mathfrak{L}}$.
{\defn
The {  graded configuration bundle} ${\pi^{\Pi\mathfrak{L}B}}$ is the triple $(\Pi\mathfrak{L}E,\pi^{\Pi\mathfrak{L}B},B)$.}

By an analogous construction as for ${\mathfrak{L}B}$, the distribution  ${\Pi\mathfrak{L}B}$ has adapted local coordinates $(x^{\alpha},u^i,\eta^a)$, where the $\eta^a$ are Grassmann-odd. Again by analogy, any element $\mathfrak{x}\varepsilon\Pi\mathfrak{L}_pE$ of the fibre at p may therefore be written in the form $\mathfrak{x}=\underset{a=1,\cdots,dim\mathfrak{G}}{\Sigma}\eta^a{\xi_a^E}$. The $\eta^a$ are uniquely determined by $(x^\alpha,u^i)$ and $\mathfrak{x}$.  They therefore determine Grassmann-odd coordinate functions $\eta:\Pi\mathfrak{L}_pE\longrightarrow\mathbb{R}^{0|dim\mathfrak{G}}$. In summary we have a coordinate chart $\mathcal{U}\times\Pi\mathfrak{L}_pE\longrightarrow\mathbb{R}^{n+m}\times\mathbb{R}^{0|dim\mathfrak{G}}\isto (p,\mathfrak{x})\longrightarrow (x^\alpha,u^i,\eta^a)$.

The graded configuration bundle  ${\pi^{\Pi\mathfrak{L}B}}$ therefore has adapted local coordinates $(x^{\alpha},u^i,\eta^a)$. The $\eta^a$ are to be identified as ``ghosts''. 
Recall that gauge symmetries in the multisymplectic formalism occur as prolongations of automorphisms of the configuration bundle \cite{Crampin}. It  is therefore  natural that the Grassmann-odd BRST symmetry ought to likewise occur as the prolongation of a Grassmann-odd bundle automorphism of the graded configuration bundle. We shall show that the BRST momentum observable is indeed the lifted momentum observable (see \S6) associated to the following Grassmann-odd globally defined vector field on ${\pi^{\Pi\mathfrak{L}B}}$, namely
\begin{equation}
V:=\eta^a\xi^{\pi^{\Pi\mathfrak{L}B}}_a -\frac{1}{2} C^c_{ab}\eta^a\eta^b\frac{\partial}{\partial\eta^c}.
\end{equation}

\subsection{The Graded Multiphase Space and Graded Configuration Bundle}
In this subsection we shall construct the graded multiphase space and the graded covariant phase space corresponding to the graded configuration bundle ${\pi^{\Pi\mathfrak{L}B}}$. We shall see that we may choose either a Grassmann-even or Grassmann-odd Cartan (n+1)-form. We shall advocate a  graded multiphase space together with the Grassmann-even Cartan (n+1)-form as the covariant multisymplectic analogue of the extended phase space of the usual BRST formalism.

The first jet bundle of ${\pi^{\Pi\mathfrak{L}B}}$, namely $\sjb$ has 1-jets which in the adapted local coordinates take the form $j^1_\mathfrak{u}\phi(x)=(x^{\alpha},u^i,\eta^a,u^i_{,\alpha},\eta^a_{,\alpha})$. The surjective submersion $\pi^{0 {{\Pi\mathfrak{L}B}}}$ is defined by the commutivity of the following diagram to be $\pi^{0 {{\Pi\mathfrak{L}B}}}:= \pi^{{{\Pi\mathfrak{L}B}}} \circ\pi^{0 {{\Pi\mathfrak{L}B}}}_1$, where $\pi^{0 {{\Pi\mathfrak{L}B}}}_1$ is the surjection mapping each one-jet onto its source.
\[
\xymatrix{
\sjb \ar[r]^{\pi^{0 {{\Pi\mathfrak{L}B}}}_1} \ar[d]_{\pi^{0 {{\Pi\mathfrak{L}B}}}} & \Pi\mathfrak{L}B \ar[dl]^{\pi^{{\Pi\mathfrak{L}B}}}\\
B
}
\]
Since ${\pi^{\Pi\mathfrak{L}B}}$ is a graded-manifold the forms on ${\pi^{\Pi\mathfrak{L}B}}$ are $\mathbb{Z}_2$-graded so that the bundle $\Lambda^n({\pi^{\Pi\mathfrak{L}B}})$ enjoys the following splitting property, \[\Lambda^n({\pi^{\Pi\mathfrak{L}B}}) \cong\Lambda^n({\pi^{\Pi\mathfrak{L}B}})^{\text{even}} \bigoplus  \Lambda^n({\pi^{\Pi\mathfrak{L}B}})^{\text{odd}}.\] We shall choose the graded multiphase space to have total space $\Lambda^n_1({\pi^{\Pi\mathfrak{L}B}})^{\text{even}}$. This choice leads to the identification of a unique canonical Grassmann-even n-form on the multiphase space with the Cartan n-form.
This choice is also related to the fact that the jet-bundle and its dual are to be related by the Legendre transformation defined by a Grassmann-even Lagrangian. One may also choose a multiphase space with a Grassmann-odd Cartan form. This leads to a  bracket on the (n-1)-forms which will pair  Grassmann-odd observables to canonically conjugate Grassmann-even observables. One might call the bracket corresponding to this second choice an anti-bracket. Although it is intriguing to speculate that this second choice may be related to  the BV formulation we do not pursue this direction at this time.
The affine dual $\sdjb$ of the first jet bundle is  defined by the following short exact sequence
\[
\xymatrix{
0 \ar[r] & \Lambda^n_0({\pi^{\Pi\mathfrak{L}B}})^{\text{even}} \ar[rr]^i&& \Lambda^n_1({\pi^{\Pi\mathfrak{L}B}})^{\text{even}} \ar[rr]^{\varrho^{\Pi\mathfrak{L}B}} & &\sdjb \ar[r] & 0
}
\]
$\Lambda^n_0({\pi^{\Pi\mathfrak{L}B}})$ and $\Lambda^n_1({\pi^{\Pi\mathfrak{L}B}})$ are defined in the same manner as in the ungraded context (see \S6). Let $\smsp$ denote $\Lambda^n_1({\pi^{\Pi\mathfrak{L}B}})^{\text{even}}$. Then $\mathcal{S}\pi$ has adapted local coordinates $(x^{\alpha},u^i,\eta^a,p^{\alpha}_i,\mathcal{P}^{\alpha}_a,p)$. The  commuting diagram below defines the  surjective submersions $\mu^{\Pi\mathfrak{L}B} := \tau^{0 \Pi\mathfrak{L}B }_1 \circ \varrho^{\Pi\mathfrak{L}B}	; \quad \kappa^{\Pi\mathfrak{L}B}:= \pi^{\Pi\mathfrak{L}B} \circ \mu^{\Pi\mathfrak{L}B}; \quad \tau^{0 \Pi\mathfrak{L}B }:=\pi^{\Pi\mathfrak{L}B} \circ \tau^{0 \Pi\mathfrak{L}B }_1$. 
\[
\xymatrix{
&\smsp \ar[ddl]_{\mu^{\Pi\mathfrak{L}B}} \ar[ddr]^{\varrho^{{\Pi\mathfrak{L}B}}} \ar[ddd]^{\kappa^{\Pi\mathfrak{L}B}} \\
&& \\
\Pi\mathfrak{L}B \ar[dr]_{\pi^{\Pi\mathfrak{L}B}}  & & \sdjb \ar'[l]'[ll]_{\tau_1^{0 {\Pi\mathfrak{L}B}}} \ar[dl]^{\tau^{0 \Pi\mathfrak{L}B}}  \\
&B
}
\]
{\prop
$\smsp$ is endowed with a canonical Grassmann-even n-form $\Theta^{\Pi\mathfrak{L}B}$. In local coordinates it takes the form 
$\Theta^{\Pi\mathfrak{L}B}=pd^nx  + p^{\alpha}_idu^i \wedge d^{n-1}x_{\alpha} +\mathcal{P}^{\alpha}_ad\eta^a \wedge d^{n-1}x_{\alpha}$.}
\begin{proof} The statement follows from the tautology $\Tilde{\omega}^*\Theta^{\Pi\mathfrak{L}B}={\omega}$ where $\Tilde{\omega}$ is the section corresponding to the Grassmann-even n-form $\omega$ at $(x^{\alpha},u^i,\eta^a)$ of $\pi^{\Pi\mathfrak{L}B}$ which corresponds to point $(x^{\alpha},u^i,\eta^a,p^{\alpha}_i,\mathcal{P}^{\alpha}_a,p)$ of $\smsp$. In local coordinates $\omega$ takes the form
$\omega=pd^nx  + p^{\alpha}_idu^i \wedge d^{n-1}x_{\alpha} +\mathcal{P}^{\alpha}_ad\eta^a \wedge d^{n-1}x_{\alpha}$.
\end{proof}
{\defn The pair $(\smsp,\Omega^{\Pi\mathfrak{L}B})$ is the {\it graded multiphase space}, where $\Omega^{\Pi\mathfrak{L}B}:=-d\Theta^{\Pi\mathfrak{L}B}$.}

In the local adapted coordinates on $\smsp$, namely $(x^{\alpha},u^i,\eta^a,p^{\alpha}_i,\mathcal{P}^{\alpha}_a,p)$, the Cartan (n+1)-form takes the form 
\begin{equation}\Omega^{\Pi\mathfrak{L}B}=-dp\wedge d^nx  + du^i \wedge dp^{\alpha}_i \wedge d^{n-1}x_{\alpha} -d\eta^a \wedge d\mathcal{P}^{\alpha}_a\wedge d^{n-1}x_{\alpha}
\end{equation}

{\defn
A {  Hamiltonian} $\mathfrak{H}$ on $\sdjb$ is a (local) section of the line bundle $(\mathcal{S}\pi,\varrho^{\Pi\mathfrak{L}B},\sdjb)$.
}
{\defn The pair $(\sdjb,\Omega^{\Pi\mathfrak{L}B}_\mathfrak{H})$ is the {  covariant phase space} where the {\it Cartan form} on ${\Omega^{\Pi\mathfrak{L}B}_\mathfrak{H}}$ is obtained from that on $\mathcal{S}\pi$ by pulling it back via the section $\mathfrak{H}$. The pair $(\sdjb,\Omega^{\Pi\mathfrak{L}B}_\mathfrak{H})$ is a graded multisymplectic manifold.
}
{\defn
Let  $\Tilde{j^1\phi} \varepsilon \Gamma(B,\sdjb)$  be the prolongation to $(\sdjb,\Omega^{\Pi\mathfrak{L}B}_\mathfrak{H})$ of a section $\phi$ of the graded configuration bundle. $\Tilde{j^1\phi}$ satisfies Hamilton's equations iff $\Tilde{j^1\phi}^*(\xi \hook \Omega^{\Pi\mathfrak{L}B}_\mathfrak{H}))=0$
$\forall \xi \varepsilon \Gamma(\sdjb,T\sdjb)$
}

For a (local) section $\mathfrak{H}$ of the line bundle $(\mathcal{S}\pi,\varrho^{\Pi\mathfrak{L}B},\sdjb)$ given by $p=-\mathfrak{H}(x^{\alpha},u^i,\eta^a,p^{\alpha}_i,\mathcal{P}^{\alpha}_a)$ the covariant Hamiltonian equations take the form
\begin{equation}
\begin{split}
&\frac {\partial p^{\alpha}_i} {\partial x^{\alpha}} =- 
\frac {\partial \mathfrak{H}} {\partial u^i}
\hspace{20pt}
\frac {\partial u^i} {\partial x^{\alpha}} = 
\frac {\partial \mathfrak{H}} {\partial p^{\alpha}_i}\\
&\frac {\partial \mathcal{P}^{\alpha}_a} {\partial x^{\alpha}} =- 
\frac {\partial \mathfrak{H}} {\partial \eta^a}
\hspace{20pt}
\frac {\partial \eta^a} {\partial x^{\alpha}} = -
\frac {\partial \mathfrak{H}} {\partial \mathcal{P}^{\alpha}_a} 
\end{split}
\end{equation}

\section{The Geometric Multisymplectic BRST Formulation of Marsden-Weinstein Reduction for first order field theories}
In this section we  translate the algebraic 
structures of the homological description of Marsden-Weinstein multisymplectic reduction \cite{jdg1} into  geometric structures on graded-multisymplectic manifolds. We identify the differential complex of \cite{jdg1} with a pair consisting of a subalgebra of the  algebra of observables on $\sdjbh$ and a  Grassmann-odd (n-1)-form. It then follows that the observables on the reduced multisymplectic manifold are obtained as the zeroth homology of this geometric differential complex.
\subsection{The $\mathbb{Z}_2\times\mathbb{Z}_2$-graded Poisson-Leibniz Algebra of Observables}
In this subsection we shall describe the algebra of observables corresponding to the graded covariant phase space $\sdjbh$.
The analogue of the $\mathbb{Z}_2$-graded Poisson-Leibniz algebra of observables on a multisymplectic manifold \cite{jdg1,kan1,kan2,kan3} is a bi-graded Poisson-Leibniz Algebra.
{\defn The {  algebra of observables} on $\sdjbh$ is the $\mathbb{Z}_2\times\mathbb{Z}_2$-graded Poisson-Leibniz algebra $(\Lambda^*(\sdhftv),\{\cdot,\cdot\})$. Any element $\mathfrak{K}\varepsilon\Lambda^*(\sdhftv)$ has bi-degree $(g_\mathfrak{K},h_\mathfrak{K})$  where $g_\mathfrak{K}=n-|\mathfrak{K}|-1$, and $h_\mathfrak{K}$ is equal to 1 if $\mathfrak{K}$ is Grassmann-odd and is equal to 0 if $\mathfrak{K}$ is Grassmann-even ($|\mathfrak{K}|$ is the form degree of $\mathfrak{K}$). The {  Poisson-Leibniz bracket} is defined to be $\{\mathfrak{F},\mathfrak{G}\}:=(-1)^{n-|\mathfrak{F}|}\Check{X}_\mathfrak{F}\hook d\mathfrak{G}$ and satisfies the following properties:
1. The {\it  left bi-graded Loday identity}\newline
\begin{equation}\{\{\mathfrak{F},\mathfrak{G}\},\mathfrak{H}\}=\{\mathfrak{F},\{\mathfrak{G},\mathfrak{H}\}\}-(-1)^{g_\mathfrak{F} g_\mathfrak{G}}(-)^{h_\mathfrak{F}h_\mathfrak{G}}\{\mathfrak{G},\{\mathfrak{F},\mathfrak{H}\}\}
\end{equation}
2. The {\it  bi-graded right Poisson-Leibniz rule}
\begin{equation}
\{\mathfrak{F}\wedge \mathfrak{G},\mathfrak{H}\}=\mathfrak{F}\wedge\{\mathfrak{G},\mathfrak{H}\} +(-1)^{|\mathfrak{G}|(g_\mathfrak{H})}(-)^{h_\mathfrak{H}h_\mathfrak{G}}\{\mathfrak{F},\mathfrak{H}\}\wedge \mathfrak{G}
\end{equation}
}
As a consequence of the left bi-graded Loday identity we have the following {\it  generalised bi-graded commutivity} (c.f. \cite{kan2}), namely
\begin{equation}\{\{\mathfrak{F},\mathfrak{G}\},\mathfrak{H}\}=-(-)^{g_\mathfrak{F}g_\mathfrak{G}}(-)^{h_\mathfrak{F}h_\mathfrak{G}}\{\{\mathfrak{G},\mathfrak{F}\},\mathfrak{H}\}
\end{equation}
We have the following subalgebras of $(\Lambda^*(\sdhftv),\{\cdot,\cdot\})$:
\begin{enumerate}
\item (i) The
$\mathbb{Z}_2\times\mathbb{Z}_2$-graded {  Lie algebra of Hamiltonian
forms} on \newline $\sdjbh$, namely $(\sdhftv,\{\cdot,\cdot\})$, and  
\item (ii) The  $\mathbb{Z}_2$-graded {  Lie algebra of Hamiltonian (n-1)-forms} on \newline $\sdjbh$, namely $(\sdhftv,\{\cdot,\cdot\})$.
\end{enumerate}

\subsection{The Origin of the BRST Symmetry}
In this subsection we shall detail the form that the classical BRST symmetry takes within the multisymplectic formalism. More precisely we shall show that the BRST multimomentum observable arises by virtue of the prolongation to the graded configuration phase space of a Grassmann-odd, commuting automorphism of the graded configuration bundle.

Recall that the sections of the distribution $\Pi\mathfrak{L}B$ take the form \begin{equation}\mathfrak{x}=\underset{a=1,\cdots,dim\mathfrak{G}}{\Sigma}\eta^a\xi^{\Pi\mathfrak{L}B}_a.\end{equation} 
The Lie bracket of two such sections as vector fields on  $\pi^{\Pi\mathfrak{L}B}$ is then \begin{equation*}[\mathfrak{x},\mathfrak{x}]=\underset{a,b=1,\cdots,dim\mathfrak{G}}{\Sigma}\eta^a\eta^bC_{ab}^c\xi^{\Pi\mathfrak{L}B}_c.\end{equation*}  In \S{\it{2.1}} we introduced a globally defined Grassmann-odd vector field on $\pi^{\Pi\mathfrak{L}B}$ which is constructed by adding to a typical section of $\Pi\mathfrak{L}B$ a term involving the structure constants of the group action, namely $V:=\eta^a\xi^{\Pi\mathfrak{L}B}_a -\frac{1}{2} C^c_{ab}\eta^a\eta^b\frac{\partial}{\partial\eta^c}$. That this vector field is indeed globally defined is guaranteed by the triviality of Lagrange multiplier distribution. Adding the extra term has the effect that this Grassmann-odd vector field has vanishing Lie bracket with itself, that is $[V,V]=0$. The vector field  generates a Grassmann-odd  automorphism of  the graded configuration bundle. So corresponding to the (possibly non-Abelian) group action on $E$ by bundle automorphisms we have constructed a Grassmann-odd commuting bundle automorphism of $\pi^{\Pi\mathfrak{L}B}$. This is the  origin of the classical BRST symmetry. The essence of the BRST symmetry is thus an ``abelianisation'' of bundle automorphisms. One thus trades a non-Abelian Grassmann-even bundle morphism for an Abelian Grassmann-odd bundle morphism.

The lifted momentum observable corresponding to the vector field $V$ is defined by $\Upsilon:=(\tau^{ 0 \Pi\mathfrak{L}B}_1)^*(V\hook \omega)$, where $\omega\varepsilon \smsp$ are n-forms on $\pi^{\Pi\mathfrak{L}B}$ (see \S6.2). In terms of the local adapted coordinates on $\smsp$,  $\omega=pd^nx  + p^{\alpha}_idu^i \wedge d^{n-1}x_{\alpha} +\mathcal{P}^{\alpha}_ad\eta^a \wedge d^{n-1}x_{\alpha}$ and the lifted momentum observable takes the form 
\begin{equation}\Upsilon :=\frac{1}{2}\underset{a,b,c=1...dim\mathfrak{G}}{\Sigma}
C^a_{b c}\eta^b\eta^c{\mathcal{P}}_a +
\underset{a=1...dim\mathfrak{G}}{\Sigma}\eta^a\delta(\xi_a).
\end{equation}
We then pull back $\Upsilon$ to $\sdjb$ via the section defined by $p=-\mathfrak{H}$ and obtain $\Upsilon_\mathfrak{H}$ which in the local adapted coordinates takes the form
\begin{equation}\Upsilon_\mathfrak{H} :=\frac{1}{2}\underset{a,b,c=1...dim\mathfrak{G}}{\Sigma}
C^a_{b c}\eta^b\eta^c{\mathcal{P}}_a +
\underset{a=1...dim\mathfrak{G}}{\Sigma}\eta^a\delta_\mathfrak{H}(\xi_a).
\end{equation}
$\Upsilon_\mathfrak{H}$ is not a Hamiltonian (n-1)-form for $\Omega^{\Pi\mathfrak{L}B}_\mathfrak{H}$ in the case where $\delta(\xi_a)$ is the lifted momentum observable corresponding to bundle automorphisms of $E$ covering non-trivial diffeomorphisms of $B$, it is however for bundle automorphisms of $E$ covering the identity on $B$.   One can most easily see that this is the case if one observes that in the former case $d\Upsilon_\mathfrak{H}$ contains a term involving $d\eta^a\wedge du^i \wedge d^{(n-2)}x_{\alpha \beta}$, which can not be obtained by taking the inner product of any vector field with $\Omega^{\Pi\mathfrak{L}B}_\mathfrak{H}$. $\Upsilon_\mathfrak{H}$ is however a generalised Hamiltonian (n-1)-form.
{\prop
The BRST n-1 form $\Upsilon_\mathfrak{H}$ generates as a  left Poisson-Leibniz derivation, namely $\{\quad,\Upsilon_\mathfrak{H}\}$ a nilpotent, generalised  multisymplectomorphism of $\sdjbh$.}
\begin{proof}
The adjoint operator  $\{\quad,\Upsilon_\mathfrak{H}\}$ is nilpotent iff $\{\Upsilon_\mathfrak{H},\Upsilon_\mathfrak{H}\}=0$. This may be seen seen from the following identity
\begin{equation}
\begin{split}
\{\{\mathfrak{F},\Upsilon_\mathfrak{H}\},\Upsilon_\mathfrak{H}\}&=\{\mathfrak{F},\{\Upsilon_\mathfrak{H},\Upsilon_\mathfrak{H}\}\} -\{\Upsilon_\mathfrak{H},\{\mathfrak{F},\Upsilon_\mathfrak{H}\}\} \text{ (Loday identity)} \\
&=\{\mathfrak{F},\{\Upsilon_\mathfrak{H},\Upsilon_\mathfrak{H}\}\} +\{\Upsilon_\mathfrak{H},\{\Upsilon_\mathfrak{H},\mathfrak{F}\}\} \text{ (generalised commutivity)}\\
&=\{\mathfrak{F},\{\Upsilon_\mathfrak{H},\Upsilon_\mathfrak{H}\}\} +\frac{1}{2}\{\{\Upsilon_\mathfrak{H},\Upsilon_\mathfrak{H}\},\mathfrak{F}\} \text{ (Loday identity)}
\end{split}
\end{equation}
Then by direct computation
\begin{equation*}
\begin{split}
d\Upsilon_\mathfrak{H} &:=d[\frac{1}{2}C^a_{b c}\eta^b\eta^c{\mathcal{P}}_a +
\eta^a\delta_\mathfrak{H}(\xi_a)] \\
	&=(-1)^nC^a_{b c}\eta^b{\mathcal{P}}_a\wedge d\eta^c +\frac{1}{2}C^a_{b c}\eta^b\eta^cd{\mathcal{P}}_a +(-)^{n-1}\delta_\mathfrak{H}(\xi_a)\wedge d\eta^a + \eta^ad\delta_\mathfrak{H}(\xi_a)
\end{split}
\end{equation*}
The {\it generalised multisymplectic structural equation} then implies that, $$\Check{X}(\Upsilon_\mathfrak{H})=(-1)^nC^a_{b c}\eta^b{\mathcal{P}}_a\circ\overset{n}{X}(\eta^c) +\frac{1}{2}C^a_{b c}\eta^b\eta^c\overset{1}{X}({\mathcal{P}}_a) +(-1)^{n-1}\delta_\mathfrak{H}(\xi_a)\circ\overset{n}{X}(\eta^a) + \eta^a X(\xi_a)$$
then,
\begin{equation*}
\begin{split}
\{\Upsilon_\mathfrak{H},\Upsilon_\mathfrak{H}\}&=\Check{X}(\Upsilon_\mathfrak{H})\hook d \Upsilon_\mathfrak{H} \\
		&=-(-1)^{2n}C^a_{b c}\eta^b{\mathcal{P}}_a\frac{1}{2}C^c_{e f}\eta^e\eta^f +\frac{1}{2}(-)C^a_{b c}\eta^b\eta^cC^g_{e a}\eta^e\eta^f{\mathcal{P}}_g \\
&-(-1)^{n-1}(-1)^{n-1}\delta_\mathfrak{H}(\xi_a)\frac{1}{2}C^a_{e f}\eta^e\eta^f
-\frac{1}{2}C^a_{b c}\eta^b\eta^c\delta_\mathfrak{H}(\xi_a) + \eta^a\eta^gC^f_{a g}\delta_\mathfrak{H}(\xi_f)
\end{split}
\end{equation*}
The first and second terms vanish by virtue of the Jacobi identity for the Lie algebra of $\mathfrak{G}$. The final three terms cancel each other so that we have obtained the desired result, namely $\{\Upsilon_\mathfrak{H},\Upsilon_\mathfrak{H}\}=0$.
\end{proof}
\subsection{The Geometric BRST Homology and The Reduced Observables}
In this subsection we shall obtain the observables on the reduced multisymplectic manifold  by virtue of a  differential complex which is constructed from natural globally defined geometric objects on the graded covariant phase space. The nilpotent differential is given by the adjoint operator $\{\quad,\Upsilon_\mathfrak{H}\}$. Recall that the BRST (n-1)-form takes the form $\Upsilon_\mathfrak{H} :=\frac{1}{2}
C^a_{b c}\eta^b\eta^c{\mathcal{P}}_a +
\eta^a\delta_\mathfrak{H}(\xi_a)$. The BRST (n-1)-form is globally defined by virtue of the triviality of the distribution $\mathcal{L}E$, for this implies that the fibre coordinates $\{\eta^a\}_{a=1,\cdots,dim\mathfrak{G}}$ and the canonical momentum observable $\{\mathcal{P}_a\}_{a=1,\cdots,dim\mathfrak{G}}$ and hence  $\Upsilon_\mathfrak{H}$ are globally defined. 

The reduced observables were obtained in \cite{jdg1} as the zeroth cohomology of the  complex $(\mathfrak{B},D)$ where $ \mathfrak{B}:=\Lambda^*(\hftv)\wedge\Lambda^*(\mathfrak{W})\otimes \Lambda^*(\mathfrak{g}^*)$. We shall show that the zeroth cohomology of this  differential complex  is isomorphic   to the Poisson-Leibniz derivation of the BRST (n-1)-form on a certain subalgebra of the Poisson-Leibniz algebra of observables on the graded covariant phase space $\sdjbh$.

Let $\mathfrak{X}$ be the vector space with basis $\{\eta^{a_1}\eta^{a_2}\cdots\eta^{a_k}\}_{a_1 <a_2 <\cdots <a_k}$ and let $\mathfrak{Y}$ be the vector space with basis $\{\mathcal{P}_{a}\}_{a=1,\cdots,dim\mathfrak{G}}$, then the exterior algebra over $\mathfrak{Y}$, namely $\Lambda^*(\mathfrak{Y})$ has the basis $\{\mathcal{P}_{a_1}\wedge\mathcal{P}_{a_2}\wedge\cdots\wedge\mathcal{P}_{a_k}\}_{a_1 <a_2 <\cdots <a_k}$. One may then form the tensor product $\Lambda^*(\mathfrak{Y})\otimes\mathfrak{X}$. We claim the algebra $\Lambda^*(\mathfrak{Y})\otimes\mathfrak{X}$ plays the same role in the geometric framework of this paper as  $\Lambda^*(\mathfrak{W})\otimes\Lambda^*(\mathfrak{g}^*)$ played in the algebraic framework of \cite{jdg1}. That $\Lambda^*(\mathfrak{g}^*)$ is graded isomorphic to $\mathfrak{X}$, where we have written  the product in the latter by juxtaposition, is clear by the correspondence of $\alpha^a\wedge \alpha^b$ to $\eta^a \eta^b$. 

When the covariant Noether currents arise by virtue of bundle automorphisms of $E$ covering the identity on $B$ (e.g. for background field theories), we claim that $\Lambda^*(\mathfrak{Y})$ and $\Lambda^*(\mathfrak{W})$ are isomorphic as graded associative algebras. In the more general case of  non-trivial diffeomorphisms of $B$ (e.g. for parametrised field theories), we shall see that $\Lambda^p(\mathfrak{Y})$ is isomorphic to  $\Lambda^p(\mathfrak{W})$ for all p such that $p(n-1)<n$. In particular for p=0,1 they are always isomorphic.

That $\mathfrak{Y}$ and $\mathfrak{W}$ are isomorphic as vector spaces follows from their definitions as  $\text{dim}\mathfrak{W}=\text{dim}\mathfrak{Y}=\text{dim}\mathfrak{G}$. The graded associative products on $\Lambda^*(\mathfrak{W})$ and $\Lambda^*(\mathfrak{Y})$ are identical. To see this  recall that (see \cite{jdg1}) for $\mathfrak{w}\varepsilon\Lambda^r(\mathfrak{W})$ and $\Acute{\mathfrak{w}}\varepsilon\Lambda^s(\mathfrak{W})$ one has $\mathfrak{w}\wedge\Acute{\mathfrak{y}}=(-1)^{rs(n^2)}\Acute{\mathfrak{w}}\wedge\mathfrak{w}$, then for $\mathfrak{y}\varepsilon\Lambda^r(\mathfrak{V})$ and $\Acute{\mathfrak{y}}\varepsilon\Lambda^s(\mathfrak{Y})$ one has $\mathfrak{y}\wedge\Acute{\mathfrak{y}}=(-1)^{rs(n-1)^2}(-1)^{rs}\Acute{\mathfrak{y}}\wedge\mathfrak{y}=(-1)^{rs(n^2)-2rsn}\Acute{\mathfrak{y}}\wedge\mathfrak{y}=(-1)^{rs(n^2)}\Acute{\mathfrak{y}}\wedge\mathfrak{y}$ as required. The structure of the basis defining $\mathfrak{W}$  is such that $\Lambda^s(\mathfrak{W})\cong \varnothing$ for all $s\varepsilon\mathbb{Z}^+$ such that $s(n-1)>n$ for the case of automorphisms of $E$ covering the identity on $B$ and $\Lambda^s(\mathfrak{W})\cong \varnothing$ for all $s\varepsilon\mathbb{Z}^+$ such that $s(n-2)>n$ for the case of automorphisms of $E$ covering nontrivial diffeomorphisms  on $B$. Correspondingly $\Lambda^s(\mathfrak{Y})\cong \varnothing$ for all $s\varepsilon\mathbb{Z}^+$ such that $s(n-1)>n$. We have therefore established that $\Lambda^*(\mathfrak{Y})\cong\underset {s=0}{\overset {s(n-1)\leq n}{\bigoplus}}\Lambda^s(\mathfrak{W})$ as graded associative algebras. 

$\djb$ is a submanifold of $\sdjb$ which is called the body of $\sdjb$. The superstructure  which arises on $\djb$ by virtue of the superstructure added to the configuration bundle to form the graded configuration bundle is called the soul of $\sdjb$ \cite{dewitt}. Since  $\djb$ is a submanifold of $\sdjb$ it follows that the graded Poisson-Leibniz algebra of observables on  $\djb$ is a Poisson-Leibniz subalgebra of the bi-graded Poisson-Leibniz algebra of observables on $\sdjb$. From $\Lambda^*(\dhftv)$ and $\Lambda^*\mathfrak{Y}\otimes\mathfrak{X}$ we may then construct the algebra $\mathfrak{A}:=\Lambda^*(\dhftv)\wedge\Lambda^*\mathfrak{Y}\otimes\mathfrak{X}$ which is the geometric  analogue of the algebra $\mathfrak{B}$ of \cite{jdg1}. 

The following theorem is the fruit  of the geometric multisymplectic formulation of the classical BRST symmetry, in which we obtain the true physical degrees of freedom of the constrained dynamical system under consideration.

{\thm The observables on the reduced multisymplectic manifold are  given by the zeroth cohomology of the complex $(\mathfrak{A},\{\quad,\Upsilon_\mathfrak{H}\})$. That is\begin{equation}
H_{\{\quad,\Upsilon_\mathfrak{H}\}}^0(\mathfrak{A})\cong\Lambda^*(\Tilde{\mathcal{H}}^*(\mathcal{B},\Omega_\mathcal{B}))
\end{equation}}
\begin{proof} (i) Let us firstly consider the case for Noether currents for which the corresponding bundle morphism is the identity on $B$. In this case we have shown that the graded vector spaces $\mathfrak{A}$ and $\mathfrak{B}$ are graded isomorphic. We wish to show that the complexes $(\mathfrak{A},\{\quad,\Upsilon_\mathfrak{H}\})$ and $(\mathfrak{B},D)$ are isomorphic. We also learn from Proposition 3.2. that $\{\quad,\Upsilon_\mathfrak{H}\}$ is nilpotent. It therefore only remains to show that the two differentials agree on generators. By direct computation one finds that this is indeed the case.
\begin{gather*}
\{\mathfrak{F},\Upsilon_\mathfrak{H}\}=\{\mathfrak{F},\delta_\mathfrak{H}(\xi_a),\} \eta^a          \qquad          [\mathfrak{F}]D=\{\mathfrak{F},\delta(\xi_a) \}\otimes \alpha^a      \\
\{\mathcal{P}_a,\Upsilon_\mathfrak{H}\}= C^d_{ab}\eta^b \mathcal{P}_d + \delta_\mathfrak{H}(\xi_a)         \qquad          [\mathfrak{w}_a]D= C^d_{ab}\mathfrak{w}_d\otimes\alpha^b + \delta(\xi_a)        \\
\{\eta^a,\Upsilon_\mathfrak{H}\}= \frac{1}{2} C^a_{bc} \eta^b \eta^c        \qquad          [\alpha^a]D=    \frac{1}{2} C^a_{bc} \alpha^b \wedge \alpha^c   
\end{gather*}
The derivation $\{\quad,\Upsilon_\mathfrak{H}\}$ then agrees on products of the generators with $D$ by virtue on the bi-graded right Leibniz rule. The derivation $\{\Upsilon_\mathfrak{H},\quad\}$ does not, since it does not satisfy a Leibniz-like rule on products. This concludes the proof as the differential complexes are isomorphic they have isomorphic cohomology and the statement follows.

(ii) For the more general case of Noether currents for which the corresponding bundle morphism involves a non-trivial diffeomorphism of $B$ the graded vector spaces $\mathfrak{A}$ and $\mathfrak{B}$ are not quite graded isomorphic. However, recall \cite{jdg1}  that the Koszul complex gives a resolution of the Poisson-Leibniz algebra of observables on the constraint submanifold. The salient feature is that the Koszul homology vanishes at all levels other than the zeroth. One thus notes that it is only $\Lambda^0(\mathfrak{W})$ and $\Lambda^1(\mathfrak{W})$ that contribute to the homology. Since $\Lambda^0(\mathfrak{W})\cong \Lambda^0(\mathfrak{Y})$ and $\Lambda^1(\mathfrak{W})\cong \Lambda^1(\mathfrak{Y})$ even in this more general case we have the desired result.
\end{proof}
\section{The Lagrange-d'Alembert Graded Formalism: Covariant Hamiltonian BFV}
The  BFV non-covariant Hamiltonian formulation of the classical BRST symmetry 
involves the introduction of the Lagrange multiplier, the ghosts, anti-ghosts and their canonically conjugate momenta. In this section we shall combine the Lagrange-d'Alembert formalism of \S1 with the graded multisymplectic formalism of \S2 in order to obtain a covariant Hamiltonian BFV formalism. 

\subsection{The Extended Lagrange Distribution}
In the Lagrange-d'Alembert formalism of \S1 we defined a certain submanifold of the graded multiphase space, by virtue of the vanishing of the Noether current $B_a$. This submanifold corresponded to  the multiphase space extended by the addition of non-propagating Lagrange multipliers. We lift the Lagrange-d'Alembert formalism off this submanifold by encoding this dynamical information into   superstructure  in the form of a certain distribution.

The Noether current $B_a$ gives rise to the action of the Abelian group $\mathfrak{K}$ of reparametrisations of the Lagrange multiplier on $\lmsp$. Let $\mathfrak{k}$ be the Lie algebra of $\mathfrak{K}$ and $\{\zeta_a\}_{a=1,\cdots,dim\mathfrak{G}}$ be a basis for $\mathfrak{k}$. The Hamiltonian vector field 
corresponding to the Hamiltonian form $J(\zeta_a):=B_a$ is $X(B_a)=\frac{\partial}{\partial \lambda^a}$. This Hamiltonian vector field pushes forward, via the surjection $\mu^{\mathfrak{L}B}$, to a  vector field corresponding to the Lie algebra element $\zeta_a$ giving rise to bundle automorphisms of the extended configuration bundle $\lcb$. The $\mathfrak{G}$ action on $\pi$  thus extendeds to a $\mathfrak{G} \times \mathfrak{K}$ action on $\lcb$.  Let $\zeta^{\lcb}_a:=\mu^{\mathfrak{L}B}_*(\frac{\partial}{\partial \lambda^a})=\frac{\partial}{\partial \lambda^a}$ and $\xi^{\lcb}_a$ denote the prolongation of $\xi^E$ from $\pi$ to $\lcb$. The $\zeta^{\lcb}_a$ and $\xi^{\lcb}_a$ span a subspace of the tangent space of $\lcb$ at $(x,u,\lambda)$. We therefore have a distribution over $\lcb$, namely 
\begin{equation}
\mathfrak{S}[\lcb]:=\underset{(x,u,\lambda) \varepsilon \lcb}{\sqcup}\mathfrak{S}[\lcb]_{(x,u,\lambda)}
\end{equation}
 where $\mathfrak{S}[\lcb]_{(x,u,\lambda)}:=<\xi^{\lcb}_a(x,u,\lambda),\zeta^{\lcb}_a(x,u,\lambda)>$. 
{\prop $\mathfrak{S}[\lcb]$ is an integrable distribution over $\lcb$.}
\begin{proof} Let $X,Y\varepsilon \mathfrak{S}[\lcb]$. In local coordinates they take the form $X=X_1^a\xi^{\lcb}_a+X_2^a\zeta^{\lcb}$ and $Y=Y_1^a\xi^{\lcb}_a+Y_2^a\zeta^{\lcb}_a$.Then
$[X,Y]=[X_1^a\xi^{\lcb}_a+X_2^a\zeta^{\lcb},Y_1^b\xi^{\lcb}_b+Y_2^b\zeta^{\lcb}_b  ]
=(X^aY^bC_{ab}^c+X^a\xi^{\lcb}_aY^c-Y^b\xi^{\lcb}_bX^c)\xi^{\lcb}_c\varepsilon\mathfrak{S}[\lcb]
$. 
\end{proof}

\subsection{The Lagrange Multiplier Extended Graded Bundles}
In this subsection we shall detail the structures of the extended graded bundles which we shall advocate as the starting point of a multisymplectic and therefore covariant Hamiltonian formulation of the classical BFV framework. We begin by defining the appropriate configuration bundle. Once we have this definition the extended graded multiphase space and the extended graded configuration phase space follow  (c.f. \S6). The diagram below illustrates the construction and defines the various surjections.
\[
\xymatrix{
\mathfrak{S}[\lcb] \ar[rr]^\Pi \ar[d]_{\pi^{\mathbb{L}}}& &\Pi\mathfrak{S}[\lcb] \ar[d]^{\pi^{\Pi\mathbb{L}}} \ar[ddl]_{\pi^{\Pi\mathbb{L}B}}\\
\mathfrak{L}E\ar[dr]_{\pi^{\mathfrak{L}B}}& &\mathfrak{L}E\ar[dl]^{\pi^{\mathfrak{L}B}} \\
&B
}
\]
{\defn{ The {  extended graded configuration bundle} $\lscb$ is the triple \newline $(\Pi\mathfrak{S}[\lcb],\lscb,B)$.}}

One can show that  one may extend the local adapted coordinates on $\pi$ to $\lscb$  which is  then  endowed with local adapted coordinates  $(x^\alpha,u^i, \lambda^a, \eta^a, \varrho^a)$ where the new coordinate $\varrho^a$ is Grassmann-odd. We call $\varrho^a$ the antighost corresponding to the Noether current $B_a$. This correspondence is made concrete by virtue of the above construction. 
The first jet bundle of $\lscb$, namely $\lsjb$ has 1-jets which in local coordinates take the form $j^1_{(u,\lambda,\eta,\varrho)}\phi(x)=(x^{\alpha},u^i,\lambda^a,\eta^a,\varrho^a,u^i_{,\alpha},\lambda^a_{,\alpha},\eta^a_{,\alpha},\varrho^a_{,\alpha})$. The surjective submersion $\pi^{0\Pi\mathbb{L}B}$ is defined by the commutivity of the following diagram to be $\pi^{0\Pi\mathbb{L}B}:=\pi^{\Pi\mathbb{L}B}\circ  \pi^{0\Pi\mathbb{L}B}_1$, where $\pi^{0\Pi\mathbb{L}B}_1$ is the surjection mapping each one-jet onto its source.
\[
\xymatrix{
\lsjb \ar[d]_{\pi^{0\Pi\mathbb{L}B}} \ar[r]^{\pi^{0\Pi\mathbb{L}B}_1}& \ar[dl]^{\pi^{\Pi\mathbb{L}B}} \Pi\mathfrak{S}[\lcb] \\
B
}
\]
Let $\Lambda^n(\lscb)^{\text{even}}$ be the subbundle of the $\text{n}^{\text{th}}$ exterior cotangent bundle over $\lscb$ whose sections are Grassmann-even n-forms on $\lscb$. The affine dual $\lsdjb$ of the first jet bundle is defined by the following short exact sequence
\[
\xymatrix{
0\ar[r] & \Lambda^n_0(\lscb)^{\text{even}} \ar[rr]^i& & \Lambda^n(\lscb)^{\text{even}} \ar[rr]^{\varrho^{\Pi\mathbb{L}B}}& & \lsdjb \ar[r] & 0
}
\]
Let $\lsmsp:=\Lambda^n(\lscb)^{\text{even}}$. $\lsmsp$ may be endowed with  local adapted coordinates $(x^{\alpha},u^i,\lambda^a,\eta^a,\varrho^a,p_i^{\alpha},B_a^{\alpha},\mathcal{P}_a^{\alpha},\mathcal{C}_a^{\alpha})$. The  commuting diagram below defines the  surjective submersions $\mu^{\mathbb{L}B} := \tau^{0 \mathbb{L}B }_1 \circ \varrho^{\Pi\mathbb{L}B}	; \quad \kappa^{\Pi\mathbb{L}B}:= \pi^{\Pi{\mathbb{L}B}} \circ \mu^{\Pi\mathbb{L}B}; \quad \tau^{0 \Pi\mathbb{L}B }:=\pi^{\Pi\mathbb{L}B} \circ \tau^{0 \Pi\mathbb{L}B }_1$. 
\[
\xymatrix{
&&\lsmsp \ar[ddll]_{\mu^{\Pi\mathbb{L}B}} \ar[ddrr]^{\varrho^{{\Pi\mathbb{L}B}}} \ar[ddd]^{\kappa^{\Pi\mathbb{L}B}} \\
&&&& \\
\Pi\mathfrak{S}[\lscb] \ar[drr]_{\pi^{\Pi\mathbb{L}B}}  &&& & \lsdjb \ar'[ll]'[llll]_{\tau_1^{0 {\Pi\mathbb{L}B}}} \ar[dll]^{\tau^{0 \Pi\mathbb{L}B}}  \\
&&B
}
\]
{\prop
$\lsmsp$ is endowed with a canonical Grassmann-even n-form $\Theta^{\Pi\mathbb{L}B}$. In local coordinates it takes the form 
$\Theta^{\Pi\mathbb{L}B}=pd^nx  + p^{\alpha}_idu^i \wedge d^{n-1}x_{\alpha} + B^{\alpha}_ad\lambda^a \wedge d^{n-1}x_{\alpha} +\mathcal{P}^{\alpha}_ad\eta^a \wedge d^{n-1}x_{\alpha} + \mathcal{C}^{\alpha}_ad\varrho^a \wedge d^{n-1}x_{\alpha}$.}
\begin{proof} The statement follows from the tautology $\Tilde{\omega}^*\Theta^{\Pi\mathbb{L}B}={\omega}$ where $\Tilde{\omega}$ is the section corresponding to the Grassmann-even  n-form $\omega$ at $(x^{\alpha},u^i,\lambda^a,\eta^a,\varrho^a)$ of $\pi^{\Pi\mathbb{L}B}$ which corresponds to point $(x^{\alpha},u^i,\lambda^a,\eta^a,\varrho^a, p^{\alpha}_i,B^{\alpha}_a,\mathcal{P}^{\alpha}_a,\mathcal{C}^{\alpha}_a,p)$ of $\lsmsp$. In local coordinates $\omega$ takes the form
$\omega=pd^nx  + p^{\alpha}_idu^i \wedge d^{n-1}x_{\alpha} + B^{\alpha}_ad\lambda^a \wedge d^{n-1}x_{\alpha} +\mathcal{P}^{\alpha}_ad\eta^a \wedge d^{n-1}x_{\alpha} + \mathcal{C}^{\alpha}_ad\varrho^a \wedge d^{n-1}x_{\alpha}$.$\square$
\end{proof}
{\defn The pair $(\lsmsp,\Omega^{\Pi\mathbb{L}B})$ is the {\it extended graded multiphase space}, where $\Omega^{\Pi\mathbb{L}B}:=-d\Theta^{\Pi\mathbb{L}B}$.}

In the local adapted coordinates on $\lsmsp$, namely $(x^{\alpha},u^i,\lambda^a,\eta^a,\varrho^a,p_i^{\alpha},B_a^{\alpha},\mathcal{P}_a^{\alpha},\mathcal{C}_a^{\alpha})$, the Cartan (n+1)-form takes the form 
\begin{equation}
\begin{split}
\Omega^{\Pi\mathbb{L}B}&=-dp\wedge d^nx  + du^i \wedge dp^{\alpha}_i \wedge d^{n-1}x_{\alpha}+ d\lambda^a \wedge dB^{\alpha}_a \wedge d^{n-1}x_{\alpha} \\
&-d\eta^a \wedge d\mathcal{P}^{\alpha}_a\wedge d^{n-1}x_{\alpha}-d\varrho^a \wedge d\mathcal{C}^{\alpha}_a\wedge d^{n-1}x_{\alpha}
\end{split}
\end{equation}

{\defn
A {  Hamiltonian} $\mathfrak{H}$ on $\lsdjb$ is a (local) section of the line bundle $(\lsmsp,\varrho^{\Pi\mathbb{L}B},\lsdjb)$.
}
{\defn The pair $(\lsdjb,\Omega^{\Pi\mathbb{L}B}_\mathfrak{H})$ is the {  covariant phase space} where the {  Cartan form} on ${\Omega^{\Pi\mathbb{L}B}_\mathfrak{H}}$ is obtained from that on $\lsmsp$ by pulling it back via the section $\mathfrak{H}$. This pair  is a graded-multisymplectic manifold.
}
{\defn
Let $\Tilde{j^1\phi} \varepsilon \Gamma(B,\lsdjb)$ be the prolongation to $\lsdjb$ of a section $\phi$ of $\lscb$. $\Tilde{j^1\phi}$ satisfies Hamilton's equations iff \[\Tilde{j^1\phi}^*(\xi \hook \Omega^{\Pi\mathbb{L}B}_\mathfrak{H}))=0\]
$\forall \xi \varepsilon \Gamma(\lsdjb,T\lsdjb)$.
}

For a (local) section $\mathfrak{H}$ of the line bundle $(\lsmsp,\varrho^{\Pi\mathbb{L}B},\lsdjb)$ given by $p=-\mathfrak{H}(x^{\alpha},u^i,\eta^a,\lambda^a,\varrho^a,p^{\alpha}_i,B^{\alpha}_a,\mathcal{P}^{\alpha}_a,\mathcal{C}^{\alpha}_a)$ the covariant Hamiltonian equations take the form
\begin{equation}
\begin{split}
&\frac {\partial p^{\alpha}_i} {\partial x^{\alpha}} =- 
\frac {\partial \mathfrak{H}} {\partial u^i} ;\hspace{10pt} \frac {\partial u^i} {\partial x^{\alpha}} = 
\frac {\partial \mathfrak{H}} {\partial p^{\alpha}_i};\\
&\frac {\partial B^{\alpha}_a} {\partial x^{\alpha}} =- 
\frac {\partial \mathfrak{H}} {\partial \lambda^a} ;\hspace{10pt} \frac {\partial \lambda^a} {\partial x^{\alpha}} = 
\frac {\partial \mathfrak{H}} {\partial B^{\alpha}_a};\\
&\frac {\partial \mathcal{P}^{\alpha}_a} {\partial x^{\alpha}} =- 
\frac {\partial \mathfrak{H}} {\partial \eta^a}; \hspace{10pt} \frac {\partial \eta^a} {\partial x^{\alpha}} = -
\frac {\partial \mathfrak{H}} {\partial \mathcal{P}^{\alpha}_a};\\
&\frac {\partial \mathcal{C}^{\alpha}_a} {\partial x^{\alpha}} =- 
\frac {\partial \mathfrak{H}} {\partial \varrho^a}; \hspace{10pt} \frac {\partial \varrho^a} {\partial x^{\alpha}} = -
\frac {\partial \mathfrak{H}} {\partial \mathcal{C}^{\alpha}_a}.
\end{split}
\end{equation}
\subsection{The Classical BRST symmetry in the Extended Graded Formalism.}
In this final subsection we describe the form that the BRST symmetry takes in the extended graded framework. The starting point is as before to notice that the group action by bundle morphisms on the extended  configuration bundle lifts to a certain Grassmann-odd bundle morphism on the extended  graded configuration bundle. The corresponding Grassmann-odd vector field commutes and in the local adapted coordinates takes the form:
\begin{equation}
V^{\pi^{\Pi\mathbb{L}B}}:=\eta^a\xi^\mathbb{E}_a -\frac{1}{2} C^c_{ab}\eta^a\eta^b\frac{\partial}{\partial\eta^c} + \varrho^a\frac{\partial}{\partial\lambda^a}
\end{equation}
The lifted momentum observable corresponding to the vector field $V^{\pi^{\Pi\mathbb{L}B}}$ is defined by $\Upsilon^{\mathbb{L}}:=(\tau^{ 1 \Pi\mathbb{L}B}_0)^*(V^{\pi^{\Pi\mathbb{L}B}}\hook \omega^{\Pi\mathbb{L}B})$, where $\omega^{\Pi\mathbb{L}B}\varepsilon \lsmsp$ are n-forms on $\pi^{\Pi\mathbb{L}B}$. In terms of the local adapted coordinates on $\lsmsp$ the lifted momentum observable takes the form 
$\Upsilon^{\mathbb{L}}:=\frac{1}{2}
C^a_{b c}\eta^b\eta^c{\mathcal{P}}_a +\eta^a\delta(\xi_a) + \varrho^aB_a$. 
We pull back $\Upsilon^{\mathbb{L}}$ by virtue of the section determined by $p=-\mathfrak{H}$ to obtain $\Upsilon^{\mathbb{L}}_\mathfrak{H}$ on $\lsdjbh$.
$\Upsilon^{\mathbb{L}}_\mathfrak{H}$ is the BRST (n-1)-form Noether current of the covariant Hamiltonian BFV formalism expounded above. By a simple calculation (c.f. Proposition 3.2.) the BRST n-1 form $\Upsilon^{\mathbb{L}}_\mathfrak{H}$ generates via its left Poisson-Leibniz action a nilpotent, generalised  multisymplectomorphism of $\lsdjbh$.

The analogue of the algebra $\mathfrak{A}$ of the graded formalism is \begin{equation}\mathfrak{A}^\mathbb{L}:=\mathfrak{A}\wedge(\Lambda^*(\mathfrak{C})\wedge\Lambda^*(\mathfrak{D}))\end{equation} 
where $\mathfrak{C}$ is the vector space with the basis $\{B_a\}_{a=1,\cdots,dim\mathfrak{G}}$ of Grassmann-even (n-1)-forms and $\mathfrak{D}$ is the vector space with basis $\{\mathcal{C}_a\}_{a=1,\cdots,dim\mathfrak{G}}$  of Grassmann-odd (n-1)-forms. The BRST homology of the extended graded formalism is defined to be $H^*_{\{\quad ,\Upsilon^\mathbb{L}_\mathfrak{H}\}}(\mathfrak{A}^\mathbb{L})$. Then by virtue of the K\"{u}nneth formula one has \begin{equation}H^p_{\{\quad ,\Upsilon^\mathbb{L}_\mathfrak{H}\}}(\mathfrak{A}^\mathbb{L})\cong \underset{m-n=p}{\bigoplus}H^m_{\{\quad ,\Upsilon_\mathfrak{H}\}}(\mathfrak{A}) \otimes H^n_{\{\quad ,\varrho^a B_a\}}(\Lambda^*(\mathfrak{C})\wedge\Lambda^*(\mathfrak{D})).\end{equation} The differential graded complex $(\Lambda^*(\mathfrak{C})\wedge\Lambda^*(\mathfrak{D}),\{\quad, \varrho^a B_a\})$ forms a Koszul complex and the homology is therefore found to be:
\begin{equation}
H^n_{\{\quad,\varrho^a B_a\}}(\Lambda^*(\mathfrak{C})\wedge\Lambda^*(\mathfrak{D}))=
\begin{cases}
0 & \text{for all }n>0 \\
\mathbb{R}& \text{ for } n=0
\end{cases}
\end{equation}
We thus learn that in particular $H^0_{\{\quad ,\Upsilon^\mathbb{L}_\mathfrak{H}\}}(\mathfrak{A}^\mathbb{L}_\mathfrak{H})\cong H^0_{\{\quad ,\Upsilon\}}(\mathfrak{A})$ and  therefore that the BRST homology in the extended graded framework is  isomorphic to the Poisson-Leibniz algebra of observables on the reduced multisymplectic manifold.

\section{An Illustration of the Covariant Hamiltonian BFV Formalism: An Application Yang-Mills Field Theory}
In this  section we shall illustrate, via application, the covariant BFV Hamiltonian formalism developed above to Yang-Mills field theory on Minkowski space-time, the principle  particle physics exemplar of a constrained classical field theory. The novelty in our covariant Hamiltonian approach is that the resulting BRST algebra is  polynomial in the canonical variables, unlike the non-covariant approach where one finds the algebra to include spatial derivatives of the canonical variables. In \cite{Mario} the authors study the BRST symmetry of the Yang-Mills theory  from the view point of the Poincar\'{e}-Cartan form on  Jet Bundles, that is in the configuration phase space approach. The presence of the derivative term in the usual BRST algebra leads them to suggest that the BRST symmetry should be formulated in a manner such that it preserves  the Poincar\'{e}-Cartan form only up to contact terms. In our approach the BRST symmetry is a  multisymplectomorphism.  

In order to present the BRST algebra in  a form close to the usual formalism we must use here Kanatchikov's formalism (\S6.3). This is because the usual  presentation of the application of the BRST formalism to examples is in terms of the Poisson bracket formulation, that is, in terms of the adjoint action of the BRST observable on the canonical coordinates and momenta. In comparing our formulation in the context of particular examples one  must therefore use the vertical covariant phase space formulation of Kanatchikov. 

We begin by defining the configuration bundle for Yang-Mills. We shall treat the connection defining the field of Yang Mills as a Lie algebra valued 1-form on Minkowski space time. The configuration bundle for Yang-Mills $\pi^{\text{YM}}$ is 
\[
\xymatrix{
\mathfrak{g} \otimes \Lambda^1{T^*\mathbb{M}} \ar[d]_{\pi^{\text{YM}}} \\
\mathbb{M} \ar@/_15pt/[u]_{\mathcal{A}_\mu^a}
}
\]
Let $E^{\text{YM}}$ denote the total space of the configuration bundle for Yang-Mills.
The first jet bundle for $\pi^{\text{YM}}$ has in local coordinates  1-jets  of the form $j^1_x\mathcal{A}(x)=(x^\mu,\mathcal{A}_\mu^a,\mathcal{A}_{\mu,\nu}^a)$. Let  $\mathcal{Z}:=\Lambda^4_1(E^{\text{YM}})$ denote the corresponding multiphase space. Then $\mathcal{Z}$  has local coordinates $(x^\mu,\mathcal{A}_\mu^a,\wp_a^{\mu \nu},p)$.
The Cartan  4-form and 5 form are then: 
\begin{equation}
\Theta^Z=pd^4x  +\wp_a^{\mu \nu} d\mathcal{A}_\mu^a \wedge d^{3}x_\nu
\end{equation}
\begin{equation}
\Omega^{\mathcal{Z}}=-dp\wedge d^4x  +d\mathcal{A}_\mu^a\wedge  d\wp_a^{\mu \nu}  \wedge d^{3}x_\nu.
\end{equation}
The pair $\mspym$ is  a multisymplectic manifold.
The dual of the first jet-bundle is defined by the following short exact sequence
\[
\xymatrix{
0 \ar[r] & \Lambda^1_0(E^{\text{YM}})\ar[r] & \mathcal{Z}\ar[r] & J^{1} \pi^{\text{YM}*}\ar[r] & 0
}
\]
Let $\vdjbym$ be the corresponding vertical configuration bundle (see \S6.3). In the local adapted coordinates the vertical Cartan 4-form has the form
\begin{equation}
\Tb:=\wp_a^{\mu \nu} d\mathcal{A}_\mu^a \wedge d^{3}x_\nu \text{ (modulo semi-basic 4-forms)}
\end{equation}
and the corresponding vertical Cartan 5-form has the form
\begin{equation}
\carb:=d\mathcal{A}_\mu^a\wedge  d\wp_a^{\mu \nu}  \wedge d^{3}x_\nu 
\end{equation}
modulo the image of semi-basic 4 forms under the exterior differential.

$\mathcal{Z}$ is too large to be the multiphase space for Yang-Mills. The appropriate multiphase space is the submanifold of $\mathcal{Z}$ identified as the image of the first jet bundle under the covariant Legendre transformation \cite{MarShk}. In order to construct the covariant Legendre transformation we will need the Yang-Mills functional $\mathcal{S}[j^1_x\mathcal{A}(x)]$, namely:
\begin{equation}
\mathcal{S}[j^1_x\mathcal{A}(x)]:=-\frac{1}{4}\int_{\mathbb{M}^4} \mathcal{F}^d_{\mu \nu}\mathcal{F}_d^{\mu \nu}
\end{equation}
where $\mathcal{F}^d_{\mu \nu}:= \mathcal{A}_{\nu,\mu}^a-\mathcal{A}_{\mu,\nu}^a + f^a_{bc}\mathcal{A}_\mu^b\mathcal{A}_\nu^c$

The covariant Legendre transformations is \cite{MarShk}:  $p^{\alpha}_i:=\frac{\partial L }{\partial v^i_{\alpha}};
p=L -\frac{\partial L }{\partial v^i_{\alpha}}(v^i_{\alpha} + \Gamma^i_{\alpha})$, where $\Gamma^i_{\alpha}$ is a connection. For Yang-Mills we define  $\Gamma^A_\nu:=:\Gamma_{\mu\hspace{5pt}\nu}^a:=-\frac{1}{2}f^a_{bc}\mathcal{A}_\mu^b\mathcal{A}_\nu^c$ where $A:=(a,\mu)_{\text{col}}$. The covariant Legendre transformation for the Yang-Mills functional is then:
\begin{equation}
\begin{align}
\wp_a^{\mu \nu} &:=\frac{\partial L }{\partial \mathcal{A}_{\mu,\nu}^a }=\mathcal{F}_d^{\mu \nu} \label{eq:CLTP}\\
p &:=L -\frac{\partial L }{\partial\mathcal{A}_{\mu,\nu}^a  } (\mathcal{A}_{\mu,\nu}^a + \Gamma_{\mu \hspace{3pt}\nu}^a) = \frac{1}{4}\mathcal{F}^d_{\mu \nu}\mathcal{F}_d^{\mu \nu} \label{eq:CLTH}
\end{align}
\end{equation}
The primary constraints, denoted  $\mathfrak{T}_a^\mu$, are determined by equation~(\ref{eq:CLTP}). Explicitly $\mathfrak{T}_a^\mu:=(\wp_a^{\mu \nu} - \mathcal{F}_a^{\mu \nu})d^{n-1}x_\nu\approx 0$. The constraints constrain the multi-momenta $\wp_a^{\mu \nu}$ to be antisymmetric  in the Greek indices, that is, $\mathfrak{T}_a^\mu \approx 0 \Longleftrightarrow \wp_a^{\mu \nu} + \wp_a^{\nu \mu} \approx 0$. This defines a submanifold of $\mathcal{Z}$ which we shall denote by $\mathcal{Z}^\mathfrak{T}$. 

The  {\it multiphase space of  Yang-Mills } is the pair
$\cmspym$, where $\carc$ is the pull back of $\Omega^{\mathcal{Z}}$ by the canonical embedding map of $\mathcal{Z}_\mathfrak{T}$ into $\mathcal{Z}$. The local adapted coordinates are $(x^\mu,\mathcal{A}_\mu^a,\mathcal{F}^{\mu\nu}_a,p)$ and the Cartan 5-form has the local coordinate expression:  
\begin{equation}
\carc:=-dp\wedge d^4x  +d\mathcal{A}_\mu^a\wedge  d\mathcal{F}_a^{\mu \nu}  \wedge d^{3}x_\nu
\end{equation}

The corresponding sub-manifold of the dual jet bundle is defined by the following short exact sequence
\[
\xymatrix{
0 \ar[r] & \Lambda^1_0(E^{\text{YM}})\ar[r] & \mathcal{Z}^\mathfrak{T}\ar[r] & J^{1} (\pi^{\text{YM}}_\mathfrak{T})\ar[r] & 0
}
\]
One may observe that the primary constraints $\mathfrak{T}_a^\mu:=(\wp_a^{\mu \nu} - \mathcal{F}_a^{\mu \nu})d^{n-1}x_\nu\approx 0$ generate an Abelian Lie algebra on $\vcdjbym$. The vertical Cartan 5-form takes the following form in terms of the local adapted coordinates
\begin{equation}
\card:=d\mathcal{A}_\mu^a\wedge  d\wp_a^{\mu \nu}  \wedge d^{3}x_\nu.
\end{equation}
Solving the multisymplectic structural equation for the Hamiltonian vector field $X(\mathfrak{T})_a^\mu$ corresponding to the momentum observable $\mathfrak{T}_a^\mu$ one finds $X(\mathfrak{T})_a^\mu=\frac{\partial}{\partial \mathcal{A}_\mu^a} +
2\eta^{\mu \lambda}\eta^{ \nu \pi}f_{abc}\mathcal{A}^c_\pi\frac{\partial}{\partial \wp_a^{\lambda \nu}}$. Finally computing the  brackets one finds:$X(\mathfrak{T})_a^\mu \hook X(\mathfrak{T})_b^\kappa \hook \card=  2\eta^{\mu \lambda}\eta^{ \nu \pi}(f_{abc}-f_{acb})\mathcal{A}^c_\pi d^{n-1}x_\nu \delta_\lambda^\kappa$. This vanishes by virtue of the  anti-symmetry of $f_{abc}$ and the algebra is thus seen to be Abelian.

The pair $\cdjbym$ is the {\it covariant phase space of Yang-Mills}, where $\care$ is the pull-back of $\carc$ by the section $  p=-\mathfrak{H}^{\text{YM}}  \varepsilon\Gamma(J^{1} \pi^{\text{YM}*},\mathcal{Z}^\mathfrak{T})$ determined by the covariant Legendre transformation (\ref{eq:CLTH}). 
The Cartan 5-form $\care$ in the local adapted coordinates on the covariant phase space is:
\begin{equation}
\begin{split}
\care&=\frac{1}{2}f^d_{ef}\mathcal{A}^e_\mu \mathcal{A}^f_\nu d\mathcal{F}_d^{\mu \nu} \wedge d^4x  -\frac{1}{2}\mathcal{F}^d_{\mu \nu} d\mathcal{F}_d^{\mu \nu}\wedge d^4x + \\ 
&\mathcal{F}_d^{\mu \nu}f^d_{ef} \mathcal{A}^e_\mu d\mathcal{A}^f_\nu \wedge d^4x  +\frac{1}{2}d\mathcal{A}_{[\mu|}^a\wedge  d\mathcal{F}_a^{\mu \nu}  \wedge d^{3}x_{|\nu]}
\end{split}
\end{equation}
Then we may determine the covariant Hamiltonian equations from
\begin{equation}
\begin{split}
\frac{\partial}{\partial\mathcal{A}^m_\pi}\hook\care=&d\mathcal{F}_m^{\pi \nu}  \wedge d^{3}x_{\nu} + \mathcal{F}_d^{\mu \pi}f^d_{ef} \mathcal{A}^e_\mu d^4x\\
\frac{\partial}{\partial\mathcal{F}_m^{\pi \kappa}}\hook\care=& \frac{1}{2}d\mathcal{A}_{[\pi|}^m  \wedge d^{3}x_{|\kappa]}+ \frac{1}{2}f^m_{ef}\mathcal{A}^e_\pi \mathcal{A}^f_\kappa d^4x
\end{split}
\end{equation}
to be 
\begin{equation}
\begin{split}
\frac{\partial\mathcal{F}_m^{\pi \kappa}}{\partial x^\kappa}+\mathcal{F}_d^{\pi \mu}f^d_{ef} \mathcal{A}^e_\mu =&0\\
\mathcal{F}^m_{\pi \kappa}=&\mathcal{A}^m_{[\pi, \kappa]} +f^m_{ef}\mathcal{A}^e_\pi \mathcal{A}^f_\kappa.
\end{split}
\end{equation}
The covariant Hamiltonian equations are thus equivalent to the source free Yang-Mills equations of motion. That is,
\begin{equation}
(\Tilde{j^1_{\mathcal{A}}})^*[U\hook\care]=0 \Longleftrightarrow
\begin{cases} 
\mathcal{F}=d_{\mathcal{A}} \mathcal{A}\\
d_{\mathcal{A}} * \mathcal{F}=0
\end{cases}
\end{equation}

We now return to the formalism of Kanatchikov (see \S6.3). 
We require that that the Hamiltonian evolution on $\vcdjbym$ should preserve the constraints. This translates into the requirement:
\begin{equation}
X(\mathfrak{T})_a^\mu\hook X(\mathfrak{H}^{\text{YM}}) \hook \card\approx 0. \label{eq:consis}
\end{equation}
The Hamiltonian $\mathfrak{H}^{\text{YM}}$ is determined by the covariant Legendre transformation (\ref{eq:CLTH}) to be $\mathfrak{H}^{\text{YM}}=-\frac{1}{4}\mathcal{F}^d_{\mu \nu}\mathcal{F}_d^{\mu \nu}$. Then
$$
\text{(\ref{eq:consis})} \Longrightarrow \mathbb{S}_a:=f_{abc}\mathcal{A}^b_\varrho \mathcal{F}^{c \varrho\nu } d^{n-1}x_\nu\approx 0
$$
We  see that the $\mathbb{S}_a$ generate the Lie algebra $\mathfrak{g}$ simply by computing  the  brackets, whence one finds that:
\begin{equation}
X(\mathbb{S})_a\hook X(\mathbb{S})_b\hook \card =f^c_{ab}\mathbb{S}_c
\end{equation}
where $X(\mathbb{S})_a$ is the solution of the multisymplectic structural equation. Explicitly:
\begin{equation}
X(\mathbb{S})_a=f_{aef}\mathcal{A}^e_\mu\frac{\partial}{\partial\mathcal{A}^f_\mu } -f_{aef}\mathcal{F}_f^{\mu \nu}\frac{\partial}{\partial\mathcal{F}_e^{\mu \nu}}
\end{equation}
By virtue of the requirement that the Hamiltonian evolution should preserve the constraints we thus find that there exits secondary first class constraints $\mathbb{S}_a:=f_{abc}\mathcal{A}^b_\varrho \mathcal{F}^{c \varrho\nu } d^{3}x_\nu$ which generate the group action by multisymplectomorphisms on $(\mathcal{Z}_A,\Omega^{\text{YM}}_A)$. This is the covariant analogue of Gauss's constraint, where unlike Gauss's constraint no derivatives of the canonical variables occur in the constraint.

Let $\smspym$ denote the {\it  extended graded multiphase space} corresponding to the configuration bundle  of  Yang-Mills. Let $\csmspym$ denote the {\it  extended graded multiphase space of Yang-Mills}  as  the submanifold of 
$\smspym$ defined by the vanishing of the primary first class constraints. 

The corresponding sub-manifold of the {\it graded extended dual jet bundle} is defined by the following short exact sequence
\[
\xymatrix{
0 \ar[r] & \Lambda^1_0(\Pi\mathcal{L}E^{\text{YM}})^*\ar[r] & \mathcal{S}^\mathbb{L}_\mathfrak{T} \ar[r] & J^{1} (\pi^{\text{YM}}_{\mathbb{L}\mathfrak{T}})^*\ar[r] & 0
}
\]

The corresponding {\it vertical covariant  phase space } is $\vsdjbym$, where the vertical Cartan 5-form $\carh$ is given in the local adapted coordinates by
\begin{equation}
\begin{split}
\carh :=& d\mathcal{A}_\mu^a\wedge  d\mathcal{F}_a^{\mu \nu}  \wedge d^{3}x_\nu
+d\lambda^a \wedge dB_a^\nu  \wedge d^{3}x_\nu \\
&-d\eta^a \wedge  d\mathcal{P}_a^{\nu}  \wedge d^{3}x_\nu
-d\varrho^a\wedge  dC_a^{\nu}  \wedge d^{3}x_\nu	
\end{split}
\end{equation}
The {\it canonical coordinate observables} on $\vsdjbym$ for Yang-Mills are:
\begin{gather*}
\mathcal{A}^a_{\mu \nu} :=\mathcal{A}^a_{\mu} d^{3}x_\nu \qquad \mathcal{F}_a^{\mu} :=\mathcal{F}_a^{\mu \nu}  d^{3}x_\nu \\
\eta^a_\nu:=\eta^a d^{3}x_\nu \qquad \mathcal{P}_a:=\mathcal{P}_a^\nu d^{3}x_\nu \\
\varrho^a_\nu:=\varrho^a d^{3}x_\nu \qquad C_a :=C_a^\nu d^{3}x_\nu \\
\lambda^a_\nu:=\lambda^a d^{3}x_\nu \qquad B_a:=B_a^\nu d^{3}x_\nu
\end{gather*}
The {\it canonical covariant  $\mathbb{Z}_2$-graded brackets} satisfied by the canonical coordinate observables are:
\begin{gather*}
\{\mathcal{A}^a_{\mu \nu} ,\mathcal{F}_b^{\kappa} \} = \frac{1}{2}\delta^a_b \delta^\kappa_{[\mu}  d^{3}x_{\nu]} \qquad \{\eta^a_\nu ,\mathcal{P}_b \} = -\delta^a_b d^{3}x_\nu \\
\{\lambda^a_\nu ,B_b \} = \delta^a_b d^{3}x_\nu \qquad \{ \varrho^a_\nu,C_b \} =- \delta^a_b d^{3}x_\nu 
\end{gather*}
One determines the Hamiltonian vector fields by solving the multisymplectic structural equation for each of the canonical coordinate observables in tern. This yields:
\begin{gather*}
X(\mathcal{A})^a_{\mu \nu} = -\frac{\partial}{\partial \mathcal{F}_a^{\mu \nu}}   ;\qquad  X(\mathcal{F})_b^{\mu}  =  \frac{\partial}{\partial \mathcal{A}^b_{\mu}}  ;      \\
X(\eta)^c_\nu  =   -\frac{\partial}{\partial\mathcal{P}_a^\nu} ;\qquad X(\mathcal{P})_a   = - \frac{\partial}{\partial \eta^a}   ;      \\
X(\lambda)^a_\nu = - \frac{\partial}{\partial B_a^\nu}  ;\qquad  X(B)_b  =  \frac{\partial}{\partial\lambda^b}   ;      \\
X(\varrho)^a_\nu = - \frac{\partial}{\partial C_b^\nu}  ;\qquad   X(C)_b =  - \frac{\partial}{\partial \varrho^b }  .     
\end{gather*}
The above Poisson brackets are then easily calculated by taking the appropriate interior products of these vector fields with $\carh$

The Hamiltonian vector field   corresponding to the BRST momentum observable, $ \Upsilon^{\text{YM}}=\eta^a\mathbb{S}_a^\nu d^{3}x_\nu -\frac{1}{2}\eta^a\eta^bf^c_{ab}\mathcal{P}^\nu_c d^{3}x_\nu +B^\nu_a\varrho^ad^{3}x_\nu$, is
\begin{equation}
\begin{split}
X(\Upsilon^{\text{YM}})=&\eta^af_{aef}\mathcal{A}^e_\mu\frac{\partial}{\partial\mathcal{A}^f_\mu }-\eta^a f_{aef}\mathcal{F}_f^{\mu \nu}\frac{\partial}{\partial\mathcal{F}_e^{\mu \nu}} +\frac{1}{2}\eta^a\eta^bf^c_{ab}\frac{\partial}{\partial \eta^c}  \\
&-(\mathbb{S}_a^\nu-\eta^bf^c_{ab}\mathcal{P}^\nu_c)\frac{\partial}{\partial\mathcal{P}^\nu_a} -B_a^\nu\frac{\partial}{\partial C_a^\nu} + \varrho^a\frac{\partial}{\partial \lambda^a}
\end{split}
\end{equation}

The BRST variations of the  canonical coordinate observables are then found to be polynomial  in the canonical variables:

\begin{equation}
\begin{split}
X(\Upsilon^{\text{YM}}) \hook X(\mathcal{A})^a_{\mu \nu} \hook \carh &=\eta^cf^a_{bc}\mathcal{A}^b_\mu d^{3}x_\nu \\
X(\Upsilon^{\text{YM}}) \hook X(\mathcal{F})_b^{\mu} \hook \carh&=-\eta^a f_{abc}\mathcal{F}_c^{\mu \nu} d^{3}x_\nu \\
X(\Upsilon^{\text{YM}}) \hook X(\eta)^c_\nu \hook \carh &=\frac{1}{2}\eta^a\eta^bf^c_{ab} d^{3}x_\nu \\
X(\Upsilon^{\text{YM}}) \hook X(\mathcal{P})_a \hook \carh &=-(\mathbb{S}_a^\nu-\eta^bf^c_{ba}\mathcal{P}^\nu_c) d^{3}x_\nu \\
X(\Upsilon^{\text{YM}}) \hook X(\lambda)^a_\nu \hook \carh &= \varrho^a d^{3}x_\nu \\
X(\Upsilon^{\text{YM}}) \hook X(\varrho)^a_\nu \hook \carh &= 0 \\
X(\Upsilon^{\text{YM}}) \hook X(C)_b \hook \carh &= -B^\nu_b d^{3}x_\nu \\
X(\Upsilon^{\text{YM}}) \hook X(B)_b \hook \carh &=  0 
\end{split}
\end{equation}

\section{Appendix: The multisymplectic formalism for first order field theories.}

In this appendix we shall briefly detail the multisymplectic framework for first order field theories. See \cite{MarShk,cramnotes,Crampin,GIMMSY} for more details. 
\subsection{The Covariant Phase Space}

The {\it fields} of a classical field theory are sections of the {\it configuration bundle} $(E,\pi,B)$ where $dimB=n$ and $dimE=m+n$.
\[
\xymatrix{
E  \ar[r]_\pi & B \ar@/_7pt/[l]_\phi
}
\]
For each $x \varepsilon B$, two sections $\phi, \psi \varepsilon \Gamma (E,B)$ have {\it first order contact at x} if:
\begin{equation}
\begin{align}
 \phi (x) &= \psi (x) \\
 \phi_{*x} (\xi) &= \psi_{*x} (\xi) \qquad \forall \xi \varepsilon T_xB.
\end{align}
\end{equation}
We denote the equivalence class of all sections with {\it{target}} $\phi(x)=u$
and {\it{source}} $x$ by $j^1_x \phi$. The set of all such equivalence classes is denoted $J_u^1 E$ and the total space of the first jet bundle is defined to be:
\begin{equation}
 \jb = \underset{u \varepsilon E}{\bigsqcup} J_u^1 E
\end{equation}
It is a bundle over E and thus also B. In fact:(i) The bundle $(\jb,\pi_1^0,E)$ is an affine bundle and (ii) The bundle $(\jb,\pi^0,B)$ is a vector bundle. If $\phi$ is a (local) section of $(E,\pi,B)$, the {\it prolongation} of $\phi$ is the section $j^1\phi$ of $(\jb,\pi_1^0,E)$ defined by $j^1\phi(x)=j^1_x\phi$. In the adapted local coordinates $j^1\phi(x)=(x^{\alpha},\phi^i,\phi^i_{,\alpha})$.
The following diagram illustrates the various surjective submersions.
\[
\xymatrix{
                           &\jb \ar[ld]_{\pi^0_1}\ar[rd]^{\pi^0}   \\
           E           \ar[rr]_\pi & &        B \\
}
\]
We now give a second characterisation of $\jb$ which we will find leads to a natural notion of a dual $\djb$. The {\it Vertical Bundle } $V \pi$ is the subbundle of the tangent bundle of $\pi$ whose fibres consist of the null space of the tangent projection $\pi_*$. The bundle $(\jb,\pi_1^0,E)$ is the affine bundle modelled on the vector bundle $V \pi \bigotimes_E T^*B$. Let $\Lambda^n E$ denote the bundle $(\Lambda^n (T^*E),\pi^*,E)$. Sections of this bundle are n-forms on $E$, $\Omega^n(E)=\Gamma(E,\Lambda^n E)$. Let $\Lambda^n_0 E$ be the bundle whose sections are in $\Omega^n_0(E)=\{\omega \varepsilon \Omega^n(E) | \xi \hook \omega =0\forall \xi \varepsilon V\pi\}$, the semi-basic n-forms, and $\Lambda^n_1 E$ be the bundle whose sections are in $\Omega^n_1(E)=\{\omega \varepsilon \Omega^n(E) | \eta \hook \xi \hook \omega =0\forall \eta, \xi \varepsilon V\pi\}$.
The Dual of $\jb$, $\djb$ is defined by the following following short exact sequence:
\[
\xymatrix{
0 \ar[r]&\Lambda^n_0 E \ar[r]& \Lambda^n_1 E \ar[r]^\varrho &\djb \ar[r]& 0
}
\]
By the identification of the affine maps from $\jb$ to $\mathbb{R}$, $\text{Aff}(\jb,\Bbb{R})$, with the space of n-forms $\Omega^n_1(E)$ and the constant affine maps from $\jb$ to $\mathbb{R}$, $\text{Aff}^{\text{const}}(\jb,\Bbb{R})$, with the space of n-forms $\Omega^n_0(E)$, the short exact sequence identifies $\djb$ as a quotient of a bundle of n-forms over E. That is, 
$\djb \cong \frac{ \text{Aff}(\jb,\Bbb{R})}{\text{Aff}^{\text{const}}(\jb,\Bbb{R})}$.

Let $z\varepsilon\Omega^n_1(E)$. Then in the local adapted bundle coordinates $z$ has the generic form $z=pd^n x + p_i^{\alpha}du_i \wedge d^nx_{\alpha}$ where $pd^n x\varepsilon\Omega^n_0(E)$ and $d^nx_{\alpha}:=\partial_{\alpha} \hook d^nx$.
One may therefore regard $(p,p_i^{\alpha})$ as local fibre coordinates on $\Lambda^n_1 E$ adapted to the fibration over $E$. The local coordinates on $(\djb,\tau^0_1,E)$ are therefore $(x^{\alpha},u^i,p_i^{\alpha})$. $\Lambda^n_1 E$ is endowed with a canonical n-form $\Theta$. In local coordinates it takes the form $\Theta=pd^n x + p_i^{\alpha}du_i \wedge d^nx_{\alpha}$. This canonical n-form is called the {\it Cartan n-form} and is an example of a multisymplectic potential. Let $\mathcal{M}\pi:=\Lambda^n_1 E$. We call the pair $(\mathcal{M}\pi,\Omega)$, where  $\Omega= -d\Theta$ the {\it Multiphase space}. The n+1-form $\Omega$ is called the {\it Cartan (n+1)-form}. $(\mathcal{M}\pi,\Omega)$ is an example of a multisymplectic manifold.
\[
\xymatrix{
           & \mathcal{M}\pi \ar[ddr]^\varrho \ar[ddl]_\mu  \ar[ddd]^\kappa \\
 & & \\
E   \ar[rd]_\pi &        & \djb \ar'[l]'[ll]_{\tau^0_1} \ar[ld]^{\tau^0} \\
           & B        
}
\]
The above commuting diagram defines the surjective submersions:
\begin{gather*}
\mu := \tau^0_1 \circ \varrho; \quad \kappa:= \pi \circ \mu; \quad \tau^0:=\pi \circ \tau^0_1.
\end{gather*}
A {\it Hamiltonian} $\mathfrak{H}$ on $\djb$ is a (local) section of the bundle $(\mathcal{M}\pi, \tau^1_0,\djb)$. The Cartan form on $\djb$ is obtained from that on $\mathcal{M}\pi$ by pulling it back via the section $\mathfrak{H}$. That is,
$\Omega_\mathfrak{H}=\mathfrak{H}^*\Omega$. The pair $(\djb,\Omega_\mathfrak{H})$, is called {\it the Covariant Phase Space}, and is a multisymplectic manifold.
A section $\Tilde{j^1\phi} \varepsilon \Gamma(\djb,B)$ satisfies Hamilton's equations if
$\Tilde{j^1\phi}^*(\xi \hook \Omega_\mathfrak{H})=0$
$\forall \xi \varepsilon \Gamma(\djb,T\djb)$.
For a (local) section  $\mathfrak{H}$ of the  bundle $(\mathcal{M}\pi, \tau^1_0,\djb)$
given by $p= -\mathfrak{H}(x^{\alpha},u^i,p^{\alpha}_i)$ the covariant Hamilton's equations take the form
\begin{equation}
\frac {\partial p^{\alpha}_i} {\partial x^{\alpha}} = -
\frac {\partial \mathfrak{H}(x^{\alpha},u^i,p^{\alpha}_i)} {\partial u^i}
\hspace{20pt}
\frac {\partial u^i} {\partial x^{\alpha}} = 
\frac {\partial \mathfrak{H}(x^{\alpha},u^i,p^{\alpha}_i)} {\partial p^{\alpha}_i}
\end{equation}

\subsection{Covariant Canonical Transformation and Hamiltonian Group Actions}

The following class of multisymplectomorphisms are the appropriate ones for studying the symmetries of field theories. All gauge symmetries arise via the prolongation of automorphism of the configuration bundle to the covariant phase space. This emphasises the role of a special class of n-1 forms called lifted momentum observables.

A {\it covariant canonical transformation}  $\Psi_{\msp}$ is a multisymplectomorphism of $(\msp,\Omega)$ such that $\Psi_{\msp}$ is a $\kappa$-bundle morphism $\Psi_{\msp}:\msp \longrightarrow \msp$ covering a diffeomorphism $\Psi_B:B\longrightarrow B$.
If $\Psi_E:E\longrightarrow E$  is a $\pi$ bundle morphism over a diffeomorphism $\Psi_B:B\longrightarrow B$ then its canonical lift to $\msp$ is given by $\Psi_{\msp}(z):={\Psi_E^{-1}}^*(z)$. $\Psi_{\msp}$ is the  prolongation the bundle morphism from $E$ to $\msp$.

Let the Lie group \group act on $\msp$ or $E$ by the bundle morphisms of $\kappa$ or $\pi$ over diffeomorphisms of $B$. Let $\Psi_B$ ,$\Psi_E$ ,$\Psi_{\msp}$ denote the transformations of $B,E,\msp$ corresponding to this group action.
 Let $\eta_B$ ,$\eta_E$ ,$\eta_{\msp}$ denote the vector fields associated with the lie algebra element $\eta \varepsilon$\lalg by virtue of this group action.

If the action $\Psi_{\msp}$ on $\msp$ is the lift of an action $\Psi_E$ on $E$, then \group acts by special covariant transformations. The {\it multimomentum map}   defined by $<\eta,\mathbb{J}(z)>:=\delta(\eta)(z):=\eta_M \hook \Omega:={{\mu}^{-1}}^*(\eta_E \hook(z))$ is a covariant momentum mapping and is $\text{Ad}^*$-equivariant. In the adapted local coordinates, the {\it lifted momentum observable}
takes the form: $\delta(\eta)(z)=(p_i^{\alpha}\eta^i +
p\eta^{\alpha})d^{n-1}x_{\alpha} - p_i^{\alpha}\eta^{\beta}du^i \wedge
\partial_{\beta} \hook (\partial_{\alpha} \hook d^n x)$ and
$\eta_E=\eta^i\frac{\partial}{\partial u^i}
+\eta^{\alpha}\frac{\partial}{\partial x^{\alpha}}$.

\subsection{Kanatchikov's ``Poisson Bracket'' Formalism and the Canonical Observables}
There is no simple canonical object on $\djb$ corresponding to the multisymplectic potential $\Theta$ on $\msp$. We have seen that we may induce multisymplectic potentials on $\djb$, determining the section by virtue of which we pull back $\Theta$ from $\msp$ to $\djb$ by the covariant Legendre transformation. An alternative would be to consider $\Theta$ modulo semi-basic n-forms, which we denote as $\Theta^V$. 

In \cite{kan1} I.Kanatchikov working with $\Theta^V$  introduces a set of canonical momentum observables. With respect to these observables he is then able to give a ``Poisson bracket'' formulation of the covariant Hamiltonian equations analogous to the familiar formulation in classical mechanics. In order to achieve this it is necessary to consider the Cartan n-form modulo semi-basic n-forms and we therefore work with 
\begin{equation}\Omega^V:= du^i \wedge dp_i^{\alpha} \wedge d^nx_{\alpha} \text{ (modulo Im$[d]\{\Omega^n_0(E)\}$). } \end{equation}

We have  therefore the multisymplectic manifold $(\djb,\Omega^V)$ which we shall call the vertical covariant phase space. One is then able to introduce canonical momentum observables $u^i_\alpha:=u^id^{n-1}x_\alpha$ and $p_i:= p_i^\alpha d^{n-1}x_\alpha$ which obey the canonical commutation relations\begin{equation} \{u^j_{\alpha} ,{p}_i \}=-=\{{p}_i ,u^j_{\alpha} \}=\delta^j_i d^{n-1}x_{\alpha}.\end{equation} Note that neither $u^i_\alpha:=u^id^{n-1}x_\alpha$ nor 
$p_i:= p_i^\alpha d^{n-1}x_\alpha$ are generically Hamiltonian forms on $(\djb,\Omega_\mathfrak{H})$. It is this fact which necessitates the introduction of $\Omega^V$.

Because the usual presentation of the application of the BRST formalism to examples is in terms of the Poisson Bracket formulation, that is, in terms of the adjoint action of the BRST observable on the canonical coordinates and momenta, in comparing our formulation in the context of particular examples one  must use the formulation of Kanatchikov. Note however that Theorem 20 is valid for the bi-graded Poisson Leibniz algebra of observables based on either $\Tilde{\mathcal{H}}^*(\djb,\Omega_\mathfrak{H})$ or $\Tilde{\mathcal{H}}^*(\djb,\Omega^V)$.

\vspace{10pt}

{\scshape{Acknowledgements}} The author would like to thank Alice Rogers for her encouragement with this article.

\end{document}